\documentclass[useAMS,usenatbib]{mn2e}
\usepackage{epsf,amsfonts,amsmath,amssymb}
\usepackage{graphicx}
\usepackage{mathrsfs}
\usepackage{color}
\usepackage{array}
\usepackage{ulem} 

\newcolumntype{C}[1]{>{\centering\let\newline\\\arraybackslash\hspace{0pt}}m{#1}}

\definecolor{blue-violet}{rgb}{0.54, 0.17, 0.89}

\title[Double Neutron Star systems Formation]{Formation of Double Neutron Star systems as implied by observations}

\author[Beniamini \& Piran ]{Paz Beniamini$^{1}$
\thanks{E-mail:paz.beniamini@gmail.com} and Tsvi Piran$^{1}$\\
$^{1}$Racah Institute for Physics, The Hebrew University, Jerusalem, 91904, Israel\\}

\begin{document}

\date{Accepted ... Received ...; in original form ...}

\pagerange{\pageref{firstpage}--\pageref{lastpage}} \pubyear{2002}

\maketitle

\label{firstpage}

\begin{abstract}
Double Neutron Stars (DNS) have to survive two supernovae and still remain bound. This sets strong limits on the nature of the second collapse in these systems.
We consider the masses and orbital parameters of the DNS population and constrain the two distributions of mass ejection and kick velocities directly from
observations with no a-priori assumptions regarding evolutionary models and/or the types of the supernovae involved. We show that there is strong evidence for two
distinct types of supernovae in these systems, where the second collapse in the majority of the observed systems 
involved small mass ejection  ($\Delta M\lesssim 0.5M_{\odot}$) and a corresponding 
low-kick velocity ($v_{k}\lesssim 30 $km\,s$^{-1}$). This formation scenario is  compatible, for example, with an electron capture supernova. Only a minority of the systems
have formed via the standard SN scenario involving larger mass ejection
of  $\sim 2.2 M_{\odot}$ and kick velocities of up to $400$km\,s$^{-1}$.
Due to the typically small kicks in most DNS (which are reflected by rather low proper motion), we predict that most of these systems reside close to the galactic disc. In particular, this implies that more
NS-NS mergers occur close to the galactic plane. This may have non-trivial implications to the estimated merger rates of DNS and to the rate of LIGO / VIRGO detections.
\end{abstract}

\section{Introduction}
\label{Int}

Double neutron star (DNS) systems are a unique and rare population of neutron stars, distinct from the population of observed isolated pulsars,
NS-WD binaries or neutron stars in X-ray binaries.  For example, the neutron stars in DNS are less heavy (and have a narrower mass distribution) than those in WD-NS systems \citep{Kiziltan(2013)}. In addition, the pulsars in these systems are rotating faster than regular
pulsars (with $P\lesssim 200$msec) and most of them are categorized as msec pulsars.
Finally the observed systems have lower proper motions than typical isolated neutron stars \citep{Lyne(1994),Hobbs(2005)}.

DNS emit gravitational radiation and consequently their orbit decays and they merge.
This make DNS systems prime candidates for detection of gravitational radiation by the upcoming advanced gravitational radiation detectors, LIGO, Virgo and KAGRA \citep{Acernese(2008),Abbott(2009),Somiya(2012)}. They are also 
prime candidates as progenitors of short GRBs \citep{Eichler(1989)} and prime candidate sources for heavy r-process nucleosynthesis \citep{Lattimer(1974),Eichler(1989)}.

A major question concerning DNS is how do the two neutron stars remain bound after the second supernova. If more than half of the mass of the system is lost, as one can expect in a hefty supernova, the system will become unbound, unless the supernova results also in a significant kick velocity to the newborn neutron star.
Assuming that the system was on a circular orbit before the second collapse (a reasonable assumption given the tidal forces between the neutron star and its companion and the high likelihood of a common envelope phase) and given the orbital parameters of the DNS system 
one can estimate what was the mass ejection and the kick velocity during the second collapse.

This method was used first by \cite{Wex(2000)} for the first binary pulsar, B1913+16, obtaining the expected velocities of a few hundred km\,s$^{-1}$ and mass ejection of a several solar masses. However, when \cite{Piran(2004)} applied this kind of analysis to the newly discovered J0737-3039 they found, surprisingly, a very strong limit on the mass ejection and kick velocity involved. Their upper limit of less then $0.1 M_\odot$,
that suggested that the second neutron star in this system was not formed in a regular supernova, was accepted with great skepticism \citep[e.g.,][]{Willems(2006)}. However this result was confirmed shortly afterwards with the strong upper limits on the proper motion of this system \citep{Kramer(2006)},  that arise naturally when a small mass is ejected \citep{Piran(2004)}. This result shows the power of this method as it gave the
first clear indication that some neutron stars (and as we argue later most in case of DNS) are not formed in a regular supernova but in an electron-capture supernova (ECS) \citep{Miyaji(1980),Nomoto(1987),Podsiadlowski(2005)}.

Currently,  13 DNS systems have been detected in the galaxy. 
There have been various attempts to estimate the mass ejection and kick velocity during the formation of the second neutron star using the orbital parameters (and sometimes pulsar spin), of the systems. These studies have focused on a single or a
few systems \citep{Wex(2000),Piran(2004),Thorsett(2005),Stairs(2006),Willems(2006),Kalogera(2007),Ferdman(2014),Dallosso(2014)}.  
Other studies carried out a population synthesis analysis for evolution of DNS systems and used the observed 
systems to constrain the parameters of the population synthesis model and obtained distributions of 
kick velocities and mass ejection. 
 \citep{Wang(2006),Wong(2010)}.

In this work we consider the present DNS population and constrain the two distributions of mass ejection and kick velocities directly from observations with no a-priori
assumptions concerning evolutionary models and/or the types of the supernovae involved. Our goal is to find mass ejection and kick velocity distributions that will explain the observed distribution of orbital parameters of the DNS population. We show that there is strong evidence for two distinct types of SNe in these systems \citep{VDH(2004),VDH(2007)},
where the majority of the observed systems have likely been formed by a low-kick velocity and low-mass ejection collapse \citep{Piran(2004)} and only a few have formed through the standard SN scenario involving large mass loss and correspondingly large kick velocities.

\section{Sample}
\label{Sample}
13 DNS systems have been detected so far in the galaxy: J0737-3039, J1906+0746, J1756-2251, B1913+16, B2127+11, B1534+12, J1829+2456, J1518+4904, J1807-2500, J1811-1736, J0453+1559, J1930-1852, J1753-2240.
In all these binaries, one of the neutron stars was detected as a pulsar. In one system,  J0737-3039,  both objects were detected as pulsars.
We exclude B2127+11 and J1807-2500 that reside in globular clusters
\citep{Jacoby(2006),Lynch(2012)} and whose orbital parameters could have changed significantly 
since their birth. We also exclude J1753-2240, for which neither the individual masses nor the total mass are well constrained.
We are thus left with 10 systems for which we obtained the values of the individual NS masses, the orbital parameters (eccentricity and orbital period), the spin period of the pulsar and either  the proper motion of the system (J0737-3039, B1534+12, B1913+16, J1518+4904)
or an upper limit on it\footnote{B1534+12 has a large measured center of mass velocity, of almost $200$km\,s$^{-1}$, this has strong implications on the derived mass ejection and kick for this system, a point we return to in detail in \S \ref{results}.} from the available literature on these systems. 
The observed parameters for all systems are summarized in Table \ref{tbl:data}.

Our goal is to constrain the mass ejection and kick velocity that take place during the formation of the second (younger) neutron star.
To this end, we first determine the orbital parameters of the system right after the explosion.
In order to do so we evolve the systems back in time taking into account losses due to gravitational radiation \citep{Peters(1964)}.
Although their spins are not as short as for the msec pulsar sub-group, most of the  observed DNS systems are fast rotators and are therefore likely to be at least mildly ``recycled", i.e. somewhat spun up by accretion from the companion. This implies that they
are the first born neutron star in their systems.  The ages of the recycled pulsars cannot reliably be estimated by the characteristic spin down times \citep{Kiziltan(2010)}. We therefore limit the time since the second collapse to be between $0 \leq t \leq t_{\rm max}$. $t_{\rm max}$ is the maximal possible age of the system, which is the minimum between the time the system's eccentricity approaches unity,
when integrating backwards in time and the Hubble time. An exception is the binary pulsar, J0737-3039, in which the second NS in the system is a regular pulsar and J1906+0746. 
This pulsar,  J1906+0746,  has the second longest spin period of the DNS systems and, more importantly, has a spin period derivative approximately four orders of magnitude larger than any of the rest and hence it is a ``regular" pulsar which does not show any indication of spin-up.
In these cases, the spin down time of the regular pulsar provides an estimate for the age of the system. 
It turns out that for most observed systems, the observed parameters have not changed significantly during their life-times (see Table \ref{tbl:data} and \S \ref{se:depend}).
Only for B1913+16 the time when the eccentricity goes to unity determines $t_{\rm max}$, and this system has a large (0.62) eccentricity also at present day.
We therefore may safely use the present day parameters in order to constrain the properties of the second collapse.
\begin{table*}
\tiny
\begin{center}
\begin{tabular}{ | C{1.3cm} | c | c | C{0.65cm} | C{0.76cm} | C{0.55cm} | C{0.6cm} | C{0.8cm} | c | c | C{0.55cm} | C{0.53cm} | C{0.75cm} | C{0.66cm} |}
\hline
System & $M_p$ ($M_{\odot}$) & $M_c$ ($M_{\odot}$) & \multicolumn{1}{|C{0.65cm}|}{\centering orbital \\ period \\ (days)} & \multicolumn{1}{|C{0.76cm}|}{\centering orbital \\ period \\ $(\text{-}t_{\rm max})$\\(days)} & $e$ & \multicolumn{1}{|p{0.8cm}|}{\centering $e$ \\$(\text{-}t_{\rm max})$ } &\multicolumn{1}{|p{0.8cm}|}{\centering pulsar \\ frequency \\ ($sec^{\text{-}1}$)}
& $\dot{p}$ & $t_c$ (Gyr) & $t_{\rm max}$ (Gyr) &\multicolumn{1}{|p{0.53cm}|}{\centering $B$ \\ ($\!10^9$G)}& \multicolumn{1}{|p{0.75cm}|}{\centering proper \\ motion \\ (km\,s$^{\text{-}1}$)} & ref. \\

\hline
J0737-3039a& $1.337^{+0.0007}_{-0.0007}$ & $1.249^{+0.0007}_{-0.0007}$ & 0.102  & 0.28 & 0.087 & 0.245 & 44 &  $1.6\! \times \!10^{-18}$ & 0.05& 0.098 & 6 & $9.6$ &\footnotemark[1]\\
J1756-2251& $1.341^{+0.007}_{-0.007}$ & $1.23^{+0.007}_{-0.007}$ & 0.319 & 0.39 & 0.181 & 0.22 &35.46 &$1.1\! \times \!10^{-18}$& 0.444 & 13.7 &5.6& $<68$ &\footnotemark[2], \footnotemark[3]\\
B1534+12  & $1.333^{+0.0002}_{-0.0002}$ & $1.345^{+0.0002}_{-0.0002}$ & 0.42 & 0.48 & 0.273 & 0.31 & 27 & $2.4\! \times \!10^{-18}$& 0.247 & 13.7 &9.5& $192$ &\footnotemark[4]\\
J1829+2456 & $1.14^{+0.28}_{-0.48}$ & $1.36^{+0.5}_{-0.17}$ & 1.17 & 1.18 & 0.139 & 0.14 & 24.4 & $5.2\! \times \!10^{-20}$& 12.4 & 13.7 &1.5& $<120$&\footnotemark[5]\\
J1518+4904  & $0.72^{+0.51}_{-0.58}$ & $2^{+0.58}_{-0.51}$ & 8.63 & 8.63 & 0.249 & 0.249 & 24.4 & $2.6\! \times \!10^{-20}$& 10.38 & 13.7 &1& $24.4$ &\footnotemark[6]\\
J0453+1559  & $1.54^{+0.006}_{-0.006}$ & $1.19^{+0.011}_{-0.011}$ & 4.07 & 4.07 & 0.11 & 0.11 & 21.8 & $1.9\! \times \!10^{-19}$& 2.63 & 13.7 &3&-&\footnotemark[7]\\
B1913+16  & $1.439^{+0.0002}_{-0.0002}$ & $1.388^{+0.0002}_{-0.0002}$ & 0.322 & 0.96 & 0.617 & 1 & 16.95 & $8.6\! \times \!10^{-18}$& 0.08 & 0.183 &23 & $75$ &\footnotemark[8]\\
J1811-1736  & $1.62^{+0.22}_{-0.55}$ & $1.11^{+0.53}_{-0.15}$ & 18.78& 18.78 & 0.828 & 0.828 & 9.61 & $9\! \times \!10^{-19}$& 1.83 & 13.7 & 9.8& $<4500$&\footnotemark[9]\\
J1906+0746& $1.291^{+0.001}_{-0.001}$ & $1.322^{+0.001}_{-0.001}$  & 0.166  & 0.29 & 0.085 & 0.15 & 6.94 &  $2\! \times \!10^{-14}$ & 0.000112 & 13.7 & 1700 & $<400$ &\footnotemark[10], \footnotemark[11]\\
J1930-1852  & $<1.32$ & $>1.3$ & 45.1 & 45.1 & 0.4 & 0.4 & 5.37 & $1.8\! \times \!10^{-17}$& 0.163 & 13.7 &59&-&\footnotemark[12]\\
\hline
\end{tabular}
 \caption{Parameters of the binary neutron star systems in the sample. $M_p$ is the pulsar mass and $M_c$ is the mass of the companion, $e$ is the eccentricity of orbit, $t_c\approx p/2\dot{p}$ is the observed pulsar's spin down
 time and $B$ is its magnetic field.}
$^1$\citealt{Kramer(2006)};
$^2$\citealt{Faulkner(2005)};
$^3$\citealt{Ferdman(2014)};
$^4$\citealt{Stairs(2002)};
$^5$\citealt{Champion(2005)};
$^6$\citealt{Janssen(2008)};
$^{7}$\citealt{Martinez(2015)};
$^8$\citealt{Weisberg(2010)};
$^{9}$\citealt{Corongiu(2007)};
$^{10}$\citealt{Lorimer(2006)};
$^{11}$\citealt{van Leeuwen(2015)};
$^{12}$\citealt{Swiggum(2015)};
\label{tbl:data}
\end{center}
\end{table*}

\section{Method}
\label{method}

We use a maximum likelihood method to estimate the most likely parameters of the distributions of mass ejection and kick velocities during the formation  of the second star in the binary.
Since the functional forms of these distributions are not known we explore various possibilities (see  \S \ref{se:depend}). We denote by $p_M(\Delta M)$ the unknown probability of a mass ejection $\Delta M$
and by $p_v(v_k)$ the unknown probability of a kick velocity $v_k$. The likelihood function (that is  maximized for the best fitting parameters of the distributions to obtain the observed parameters) is then defined by:
\begin{eqnarray}
\label{eq:likelihood}
 L(\Delta M_0,\sigma_{\Delta M},v_{k,0},\sigma_{v_k})=\Pi_i \sum_{\Delta M,v_k} p_M(\Delta M) p_v(v_k) \\  \times~ p_{\rm orb,i}(e_i,a_i | M_{p,i},M_{c,i}, v_k, \Delta M,a_{0,i},v_{\rm prop,i})
 \end{eqnarray}
where $\Pi_i$ runs over the observed DNS systems and $p_{\rm orb,i}(e_i,a_i | M_{p,i},M_{c,i}, v_k, \Delta M,a_{0,i},v_{\rm prop,i})$ is the probability of obtaining immediately after the explosion the eccentricity, $e_i$, and semi-major axis, $a_i$, for the $i$th system.
We assume that the exploding star is the companion of the observed pulsar\footnote{This is most likely the case for the majority of the systems (excluding J1906+0746) in the sample that have short spin periods and small spin period derivatives and are therefore likely recycled and born after their companion. In any case, the value of $p_{\rm orb,i}(e_i,a_i | M_{p,i},M_{c,i}, v_k, \Delta M,a_{0,i},v_{\rm prop,i})$ depends only weakly on the individual masses $M_p,M_c$.
In addition the mass difference between the two stars in DNS systems is usually small. Therefore the association of the companion with the second exploding star does not strongly affect our results.}
and that before the explosion the orbit was circular. $p_{\rm orb,i}$ is then determined by the masses of the pulsar and companion today,
$ M_{p,i}$ and $M_{c,i}$ respectively, the kick velocity magnitude $v_k$, the amount of mass ejection $\Delta M$, the semi-major axis of the system just before the explosion $a_{0,i}$ and the observed proper motion of the system $v_{\rm prop,i}$.

To calculate this probability we use the conservation of energy and angular momentum during the collapse \citep{Postnov(2014)} to relate the conditions before and after the collapse:
\begin{equation}
\label{axisratio}
\frac{a}{a_0}=\bigg[2-\chi (\frac{w_x^2+w_z^2+(v_{\rm kep}+w_y)^2}{v_{\rm kep}^2}\bigg]^{-1}
\end{equation}
and
\begin{equation}
\label{eq:eccentricity}
1-e^2=\chi \frac{a_0}{a} (\frac{w_z^2+(v_{\rm kep}+w_y)^2}{v_{\rm kep}^2}),
\end{equation}
where $\chi\equiv (M_0+M_p)/(M_c+M_p) \geq1$ is the fractional change in mass, $M_0$ is the companion mass right before the explosion, $v_{\rm kep}=\sqrt{G(M_0+M_p)/a_0}$ is the Keplerian velocity
of the initial orbit and $w_x,w_y,w_z$ are the components of the kick velocity along the axis from $M_p$ to $M_c$,
the direction of ${\vec v}_{\rm kep}$, and the axis perpendicular to the orbital plane correspondingly.
We have dropped the index $i$ appearing in Eq. \ref{eq:likelihood} for clarity and since the present equations apply to individual systems, as opposed as to the whole population.
These equations may be solved for $e,a_0$ given $a,M_c,M_p,\Delta M,w_x,w_y,w_z$.
These equations may not have a solution.  A well known example is that if there is no kick ($w_x=w_y=w_z=0$) and the system loses more than half its mass ($\chi>2$), in which case it is disrupted.  
This can be seen clearly in Eq. \ref{axisratio} which would result in a negative $a_0$ in this case.

An additional constrain on the system may be obtained if there is an estimate (or a limit) on the center of mass (CoM) velocity, $\vec{v}_{CM}$, of the system.
Using conservation of momentum one can relate the kick applied to the expelled mass to the change in $\vec{v}_{CM}$:
\begin{equation}
\label{eq:vcm}
 \Delta \vec{v}_{CM}=\frac{M_c}{M_p+M_c}\vec{v}_k+\frac{\Delta M}{M_c+M_p}\frac{M_p}{M_p+M_0} \vec{v}_{\rm kep}
\end{equation}
Since we do not know the direction of $\vec{v}_{CM}$ and its initial value, this consideration results in a softer limit on the allowed parameter space. Conservatively we impose the following condition on the observed proper motion, $\vec{v}_{\rm prop}$:
\begin{equation}
\label{eq:vcmlimits}
 \frac{\rm max(0,\Delta {v}_{CM}-50\mbox{km}\,s^{-1})}{\sqrt{1.5}}<v_{\rm prop}<\Delta {v}_{CM}+50 \mbox{km}\,s^{-1},
\end{equation}
where we have assumed that the line of sight component of the velocity is not significantly larger than the observed components: $v_{CM}<v_{\rm prop}<\sqrt{1.5} v_{CM}$ and given
typical random stellar motions in the galaxy we have taken
$50 \mbox{km}\,s^{-1}$ as a reasonable upper limit on the initial CoM velocity.
When only an upper limit on the true proper motion is available (see \S \ref{Sample}) we can only use the R.H.S. of the inequality in Eq. \ref{eq:vcmlimits}.

We use a Monte Carlo method to estimate $p_{\rm orb,i}(e_i,a_i | M_{p,i},M_{c,i}, v_k, \Delta M,a_{0,i},v_{\rm prop,i})$. For each system and for each pair $\Delta M,v_k$ we draw $10^4$ random realizations of $w_x,w_y,w_z$
(such that $w_x^2+w_y^2+w_z^2=v_k^2$ and the direction of the kick velocity is drawn from a uniform distribution on a  sphere
).
Using these values we calculate the solid angle in velocity space for which a solution with $a_i,e_i$ to Eqns. \ref{axisratio}, \ref{eq:eccentricity}, \ref{eq:vcmlimits} exists. The probability is the ratio of this solid angle to the solid angle of the whole sphere,  $\Omega/4\pi$.

\section{Results}
\label{results}

\subsection{An apparent bi-modality}
\label{sec:bimodality}
\begin{figure*}
\centering
\includegraphics[scale=0.25]{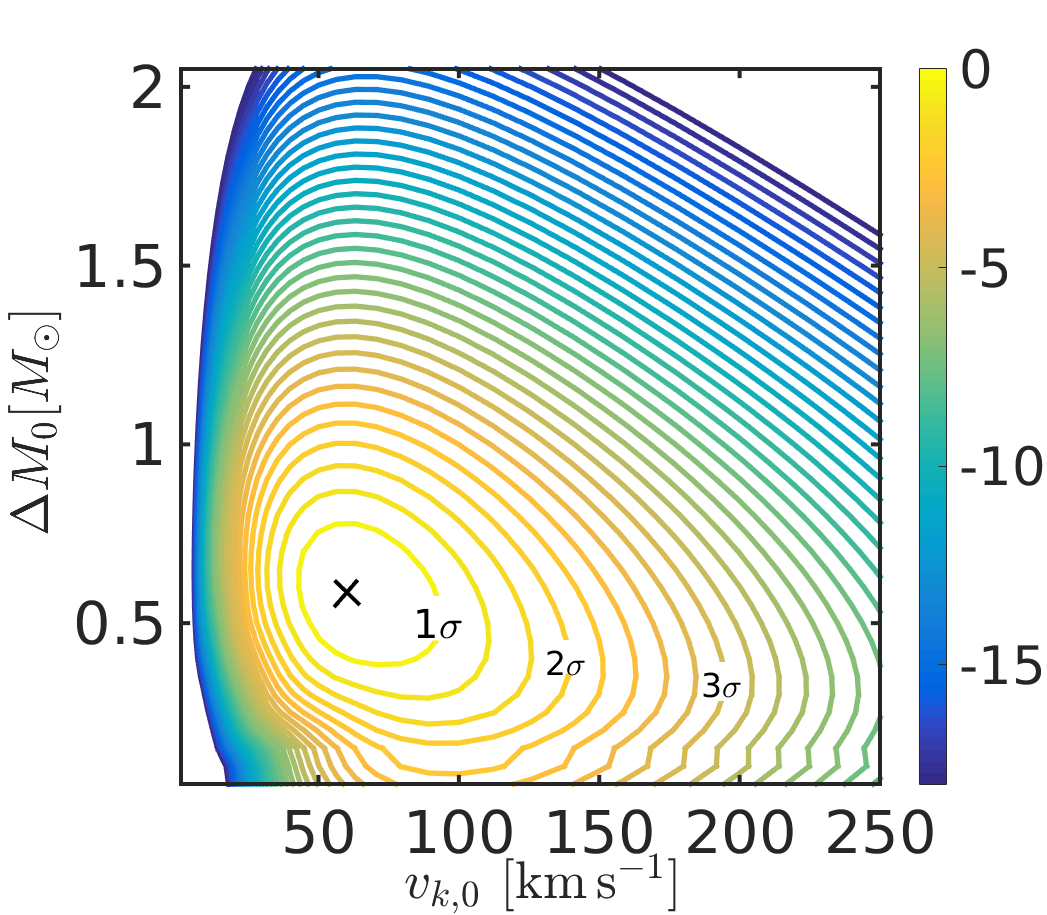}
\caption
{\small A likelihood function as a function of $\Delta M_0,v_{k,0}$ for log-normal distributions in $\Delta M,v_k$ with $\sigma_M/ \Delta M_0=\sigma_{v_k}/v_{k,0}= 0.5$ for all DNS systems in the sample. Contour lines depict 0.5 logarithmic intervals of the likelihood function relative to the maximum value. Also shown are the $1\sigma$, $2\sigma$ and $3\sigma$ confidence intervals corresponding to $\ln L=-0.5,-2,-4.5$.}
\label{fig:GausGausall}
\end{figure*}

We calculate the maximum likelihood described in \S \ref{method} using the observed orbital parameters of the systems today and assuming log-normal distributions for $\Delta M,v_k$:
\begin{equation}
\begin{array}{l}
 P(\Delta M|\Delta M_0,\sigma_{\ln \Delta M})=\frac{1}{\sqrt{2 \pi}\,\sigma_{\ln \Delta M}\Delta M}\,\exp\!\left(-\frac{[\ln(\Delta M/\Delta M_0)]^2}{2\sigma_{\ln \Delta M}^2}\right)\\
 P(v_k|v_{k,0}, \sigma_{\ln v_k})=\frac{1}{\sqrt{2 \pi}\,\sigma_{\ln v_k}v_k}\,\exp\!\left(-\frac{[\ln(v_k/v_{k,0})]^2}{2\sigma_{\ln v_k}^2}\right)
\end{array} 
\end{equation}
with $\sigma_{\ln \Delta M}=\sigma_{\ln v_k}=\sinh^{-1}(0.5)$ (corresponding to typical widths: $\sigma_M/ \Delta M_0=\sigma_{v_k}/v_{k,0}= 0.5$). We carry out this calculation, first, for the whole population of DNS systems (see Fig. \ref{fig:GausGausall}). The maximum likelihood is obtained for
$\Delta M_0=0.6M_{\odot}$ and $v_{k,0}=50 $km\,s$^{-1}$. The likelihood distribution is wide, easily accommodating values of $\Delta M_0$ between $0.15M_{\odot}$ and $1M_{\odot}$ and values of $v_{k,0}$ in the range $25-140 $km\,s$^{-1}$.
As will be shown below, this model can be ruled out. In addition there are some physical arguments that support a two population model as described in the following paragraph.

\begin{figure*}
\centering
\includegraphics[scale=0.25]{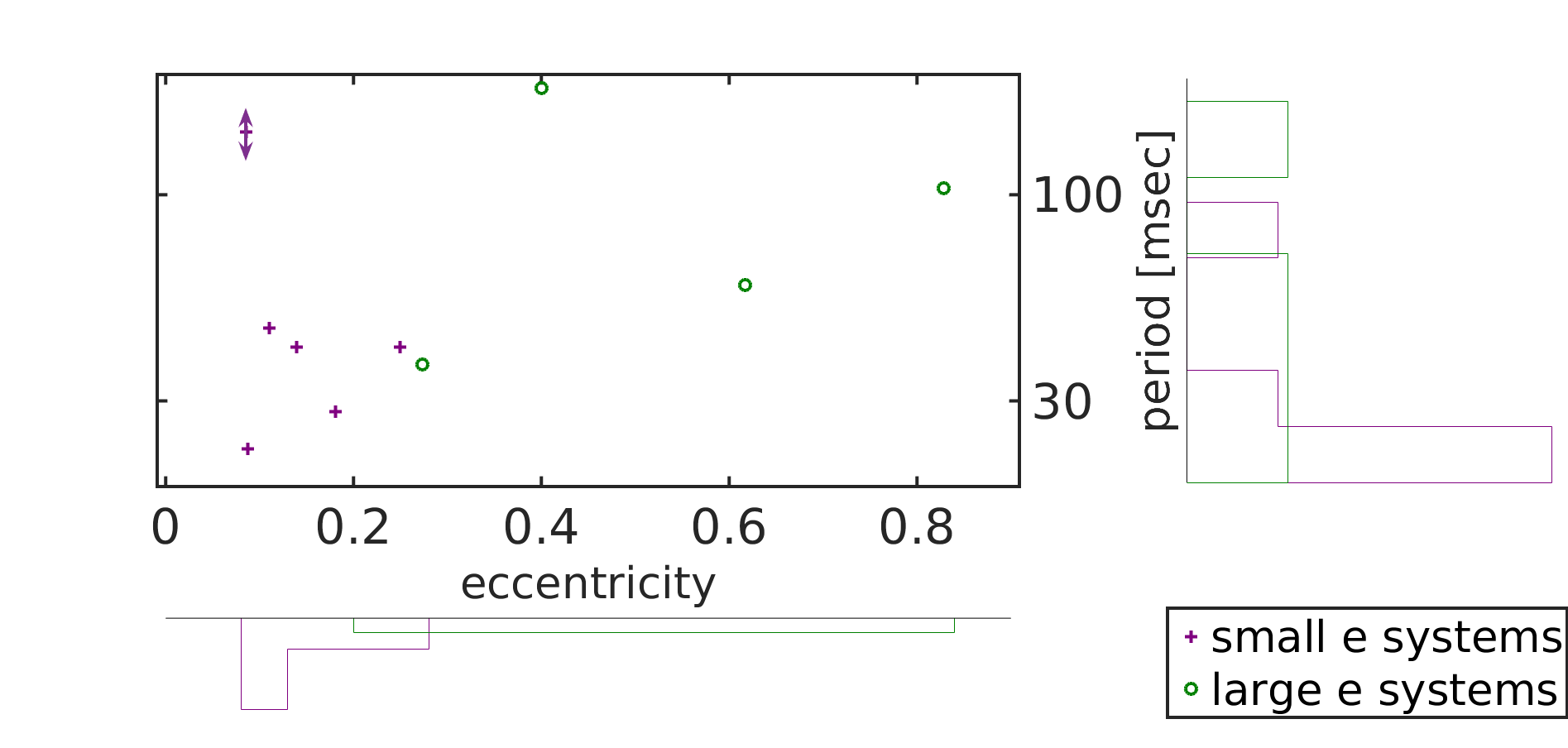}
\caption
{\small The distribution of eccentricities and spin periods of the observed pulsars in DNS systems.
An apparent bi-modality is observed between systems with small eccentricities and short pulsar periods (pluses) and those with large eccentricities and long
spin periods (circles). We denote the first as ``small $e$ systems'' and the second as ``large $e$ systems''.
In J1906+0746 (top left), the observed pulsar is a regular (as opposed to recycled) pulsar and therefore likely the second rather than the first born NS in the system. Its spin period is therefore not expected to bear any connection with
that of the other systems in the sample, and is thus marked by a double sided arrow.
A marginal case is that of B1534+12 which is on the edge of the two populations.
However, given the large center of mass velocity observed in this system and the relatively large spin tilt angle,
it likely belongs to the second group (see \S \ref{sec:bimodality} for details).}
\label{fig:elogP}
\end{figure*}

Low eccentricity indicates small mass ejection and a small kick velocity while large eccentricity indicates a large kick velocity and a significant mass ejection.
We therefore divide the system to two groups according to their eccentricities. It is interesting to note that the low eccentricity systems also typically have shorter spin periods
(see Fig. \ref{fig:elogP} and also \citealt{VDH(2007),VDH(2011)}) suggesting that indeed the two groups have physically different formation mechanisms. 
We divide the population of DNS systems to those with small eccentricities and spin periods
and those with larger spin periods and larger eccentricities\footnote{For the double binary system,
we take the period of the $22.7$msec pulsar (J0737-3039 a) rather than that of the $2.8$sec companion.}.
In what follows, we denote the first as ``small $e$ systems'' and the latter as ``large $e$ systems''.
A special case, is that of J1906+0746, which has the second largest spin period but also the smallest eccentricity ($e=0.085$). As mentioned in \S \ref{Sample}, given the extremely large value of $\dot{p}$ in this system
and the large spin period, the observed pulsar is in fact a regular pulsar and therefore likely the second rather than the first born NS in the system. Its spin period is therefore not expected to play the same role as
that of the other systems in the sample and therefore we include it in the group of low eccentricity systems. 
A somewhat marginal case is that of B1534+12 which is on the edge of the two populations.
However, given the large center of mass velocity observed in this system and the relatively large angle between the pulsar's spin and the orbital angular momentum of the binary ($\sim 25^{\circ}$),
it likely belongs to the second group which, as we shall show below, are typically fit with large mass ejections and kick velocities
(\citealt{Stairs(2006),Kalogera(2007),Wong(2010),Andrews(2015)} who discuss specifically this system reach similar conclusions).
We test the dependence of the results on this assumption in \S \ref{se:depend}, where we repeat the analysis with B1534+12 in the ``small $e$ systems" group.

\begin{figure*}
\centering
\includegraphics[scale=0.25]{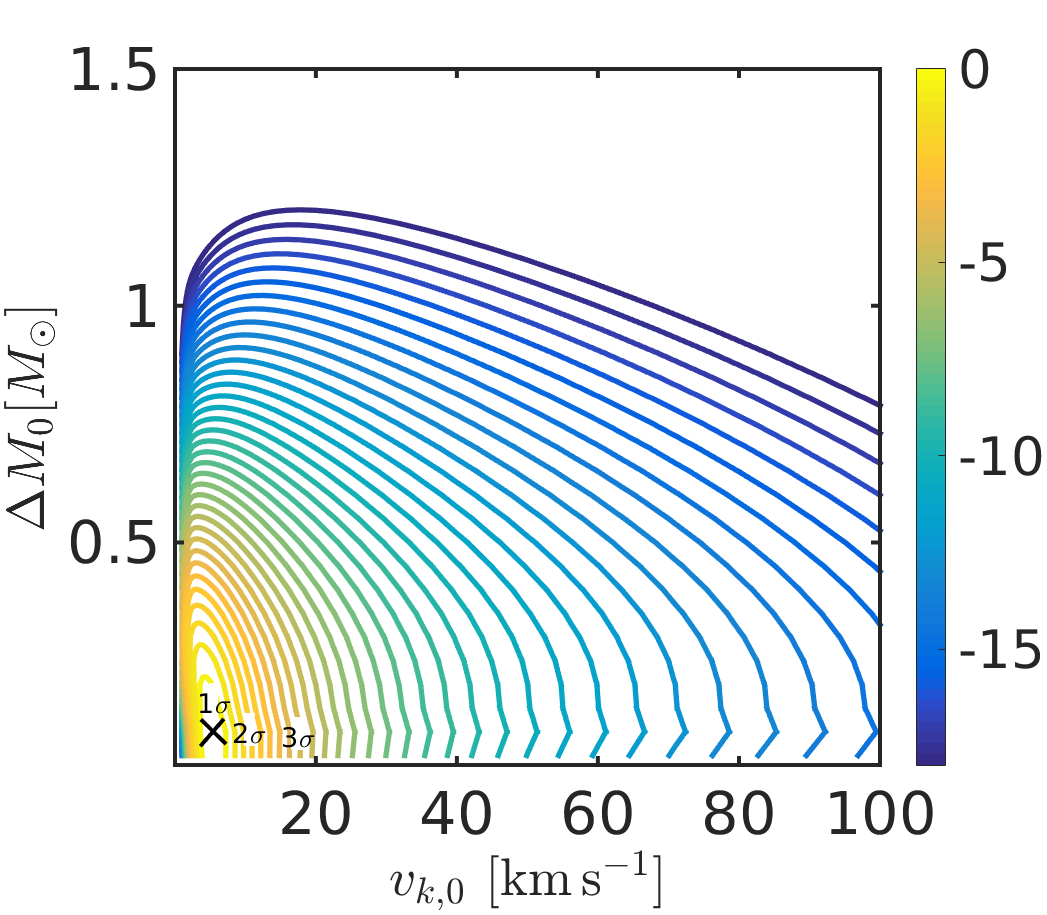}
\includegraphics[scale=0.25]{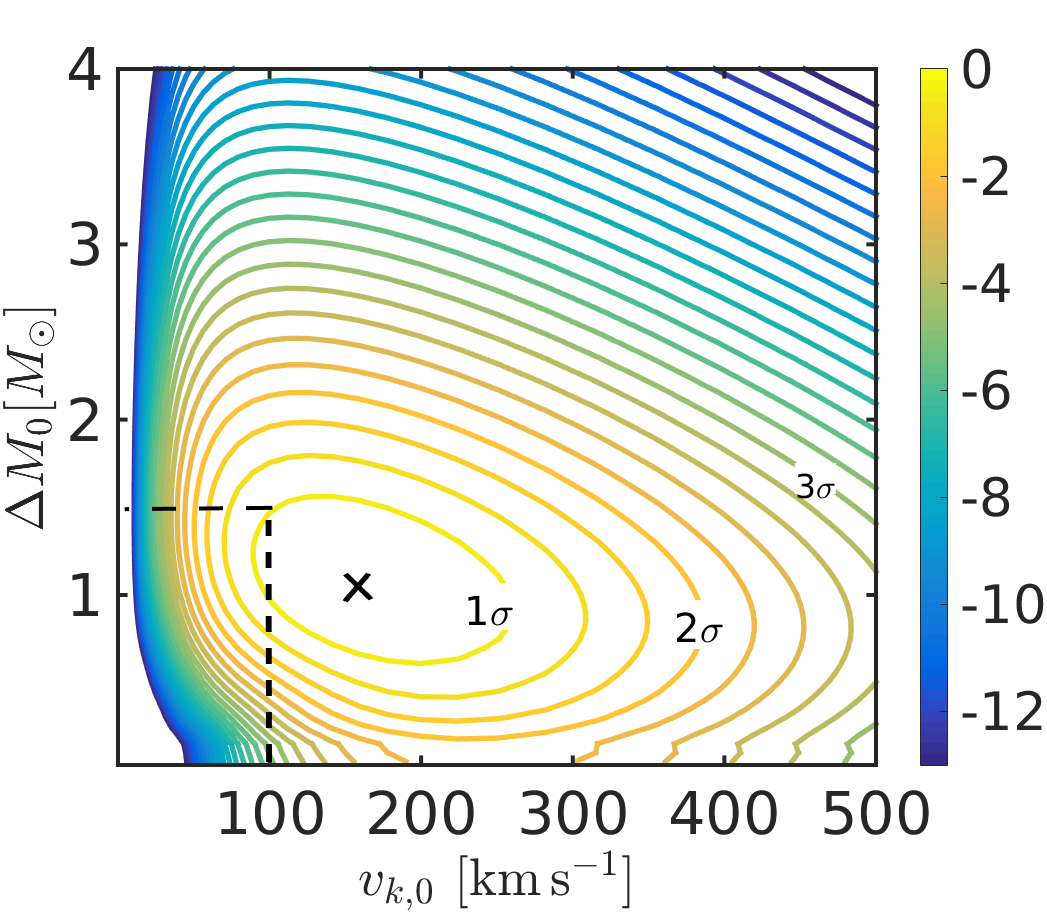}
\caption
{\small A likelihood function as a function of $\Delta M_0,v_{k,0}$ for log-normal distributions in $\Delta M,v_k$ with $\sigma_M/ \Delta M_0=\sigma_{v_k}/v_{k,0}= 0.5$. Contour lines depict 0.5 logarithmic intervals of the likelihood function relative to the maximum value. Also shown are the $1\sigma$, $2\sigma$ and $3\sigma$ confidence intervals corresponding to $\ln L=-0.5,-2,-4.5$.{\bf Left: } The 6 DNS systems
categorized as ``small $e$ systems''. {\bf Right: }The 4 DNS systems
categorized as ``large $e$ systems''. The dashed box in the right panel depicts the axis limits for the left figure.}
\label{fig:GausGaus}
\end{figure*}

\begin{figure*}
\centering
\includegraphics[scale=0.2]{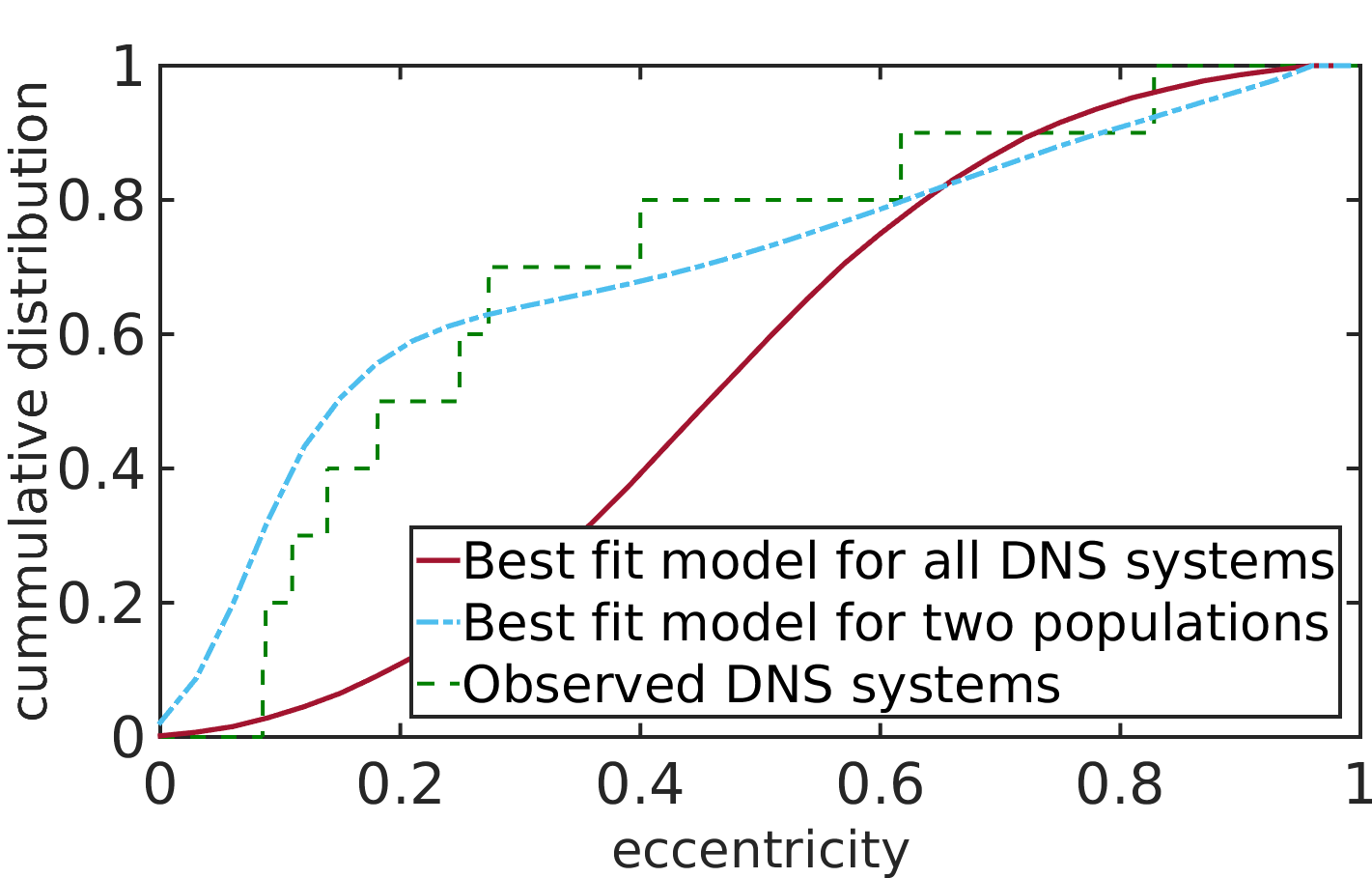}\\
\includegraphics[scale=0.2]{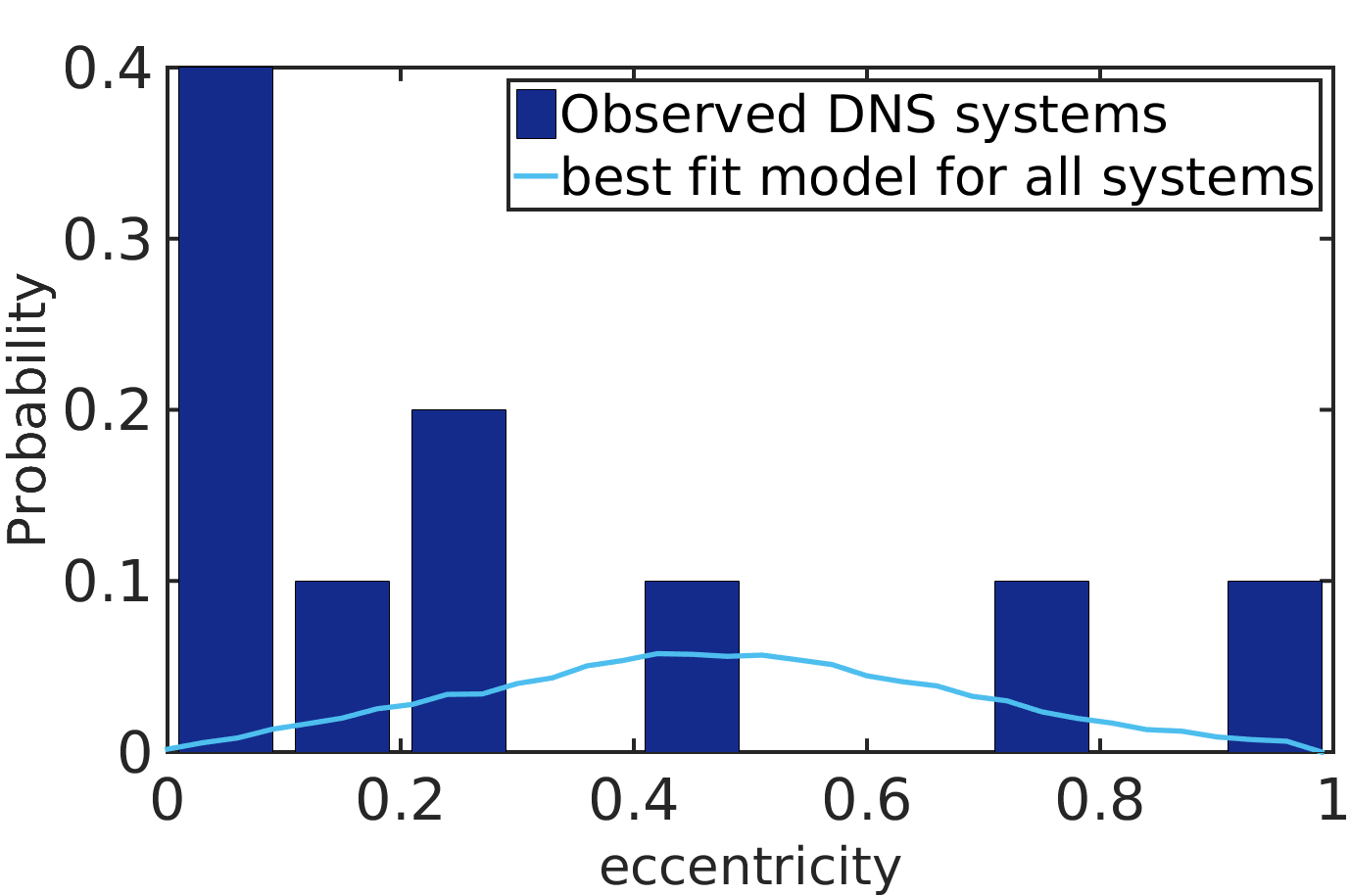}
\includegraphics[scale=0.2]{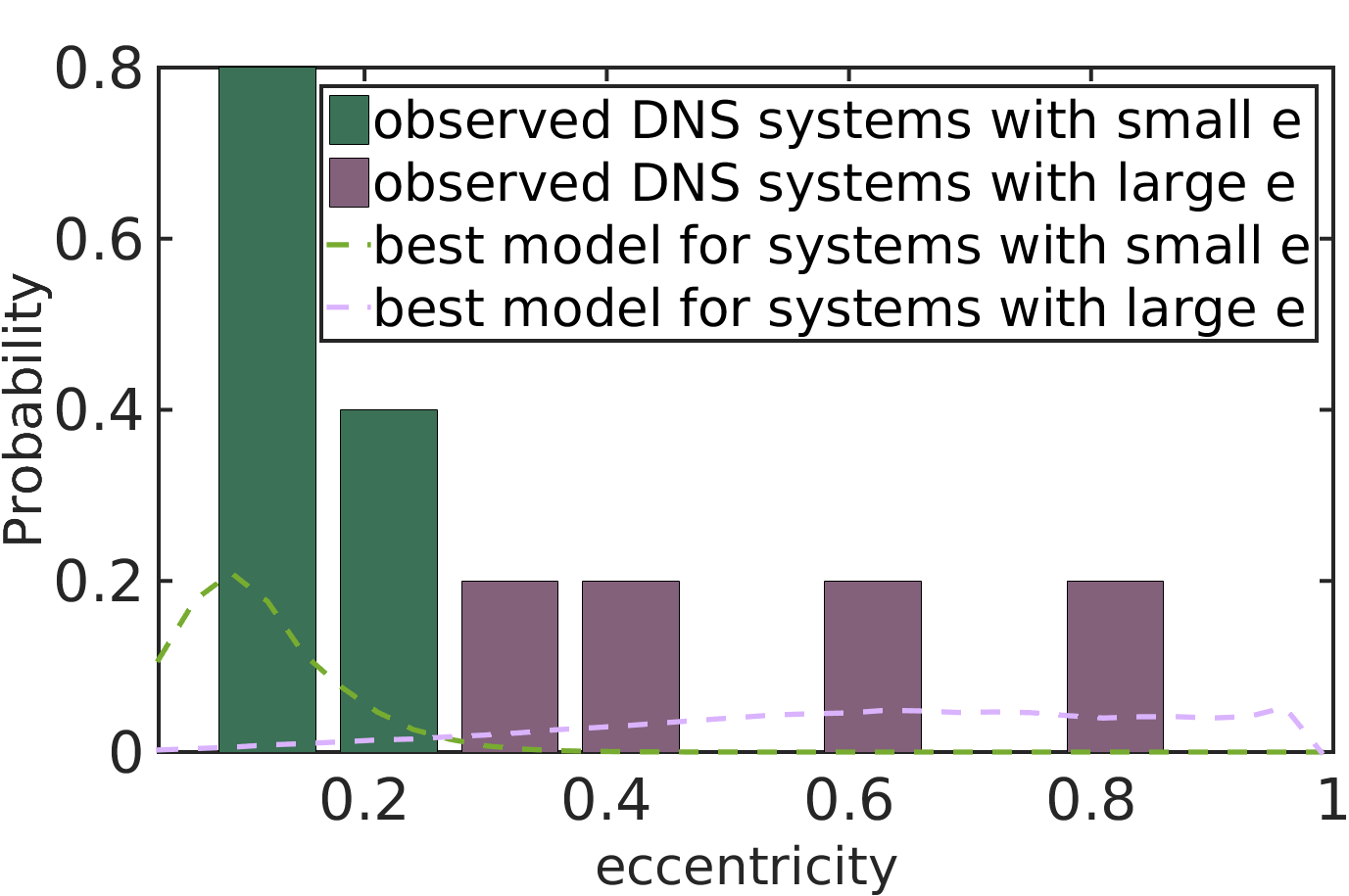}
\caption
{\small {\bf Top:} A KS test for the eccentricity distributions from the best fit model for a single DNS population (solid line) and 
for two distinct groups according to the pulsar spin periods (dot-dashed lines). The observations are depicted by the dashed line.
{\bf Bottom:} The differential distributions of observed eccentricities as compared with the best fit results for all DNS systems
(left) and for the DNS systems grouped to short and large $e$ systems (right).}
\label{fig:eProb}
\end{figure*}

We repeat the likelihood analysis, but now assuming that there are two distributions one for each group. The results are shown in Fig. \ref{fig:GausGaus}. For the ``small $e$ systems'' the likelihood peaks at
$\Delta M_0^{s}=0.1M_{\odot}$ and $v_{k,0}^{s}=5 $km\,s$^{-1}$ (with probabilities of $\Delta M_0^{s}\geq0.4$ and
$v_{K,0}^{s}>12$km\,s$^{-1}$ falling below $5\%$). These results strongly suggest an ECS-type SN as the origin of the companion in these systems.
For the ``large $e$ systems'' the likelihood peaks at $\Delta M_0^{l}=1M_{\odot}$ and $v_{k,0}^{l}=158 $km\,s$^{-1}$. In this case, the allowed range for average kick velocities is between $40-400$km\,s$^{-1}$ and average mass ejections go up to $2.2M_{\odot}$.
This suggests a more violent explosion as the origin of these systems, such as would be expected in a core-collapse SN.

In order to estimate which of the two models (a single population or two populations) better describes the observations, we preform a likelihood ratio test.
The likelihood ratio test statistic is defined by: $LR=-2\ln(L_{\rm max,1} / L_{\rm max,2})$, where $L_{\rm max,1}$ is the maximum likelihood value obtained with a single group and $L_{\rm max,2}$ is the maximum
with two separate groups. The resulting test statistic is a distributed $\chi$-squared, with degrees of freedom equal to the difference in the number of free parameters between the two models. We obtain $LR=34$ with a difference of
two degrees of freedom between the models, the associated p-value is $p<0.001$, indicating that the model with the two different groups is a significantly better descriptor of the data.
In \S \ref{se:depend} we show that this conclusion remains qualitatively unchanged for all distributions of $\Delta M$ and $v_k$ that we considered.
We also perform a KS test comparing between the observed eccentricity distributions and those predicted by the best fit model.
The KS test rules out the single population model at the $99\%$ confidence level, whereas the two populations model is consistent (see top left panel of Fig. \ref{fig:eProb}).

The need to divide the population of DNS to two groups can be understood intuitively in terms of the differential eccentricity distribution.
As mentioned above, there is a significant group of DNS systems with a narrow distribution of small eccentricities ($0.1 \lesssim e \lesssim 0.25$), while other systems have a much wider range of eccentricities.
For large kick velocities ($v_k \gg v_{\rm kep}$) one can expect a wide distribution in eccentricities, peaking at increasingly larger eccentricities as the kick velocity increases.
However, for sufficiently low kick velocities the eccentricity can be approximated by: $e \approx \Delta M/ M_{tot}$ \citep{BVDH(1991)}. The observed distribution of eccentricities, composed of systems with short pulsar periods, all
concentrated within a narrow range of eccentricities and systems with longer pulsar periods and a wider distribution of eccentricities, makes the need for the two intrinsically different populations, intuitively clear
(see bottom panels of Fig. \ref{fig:eProb}).

\subsection{Dependence on the model parameters}
\label{se:depend}
\begin{figure*}
\centering
\includegraphics[scale=0.24]{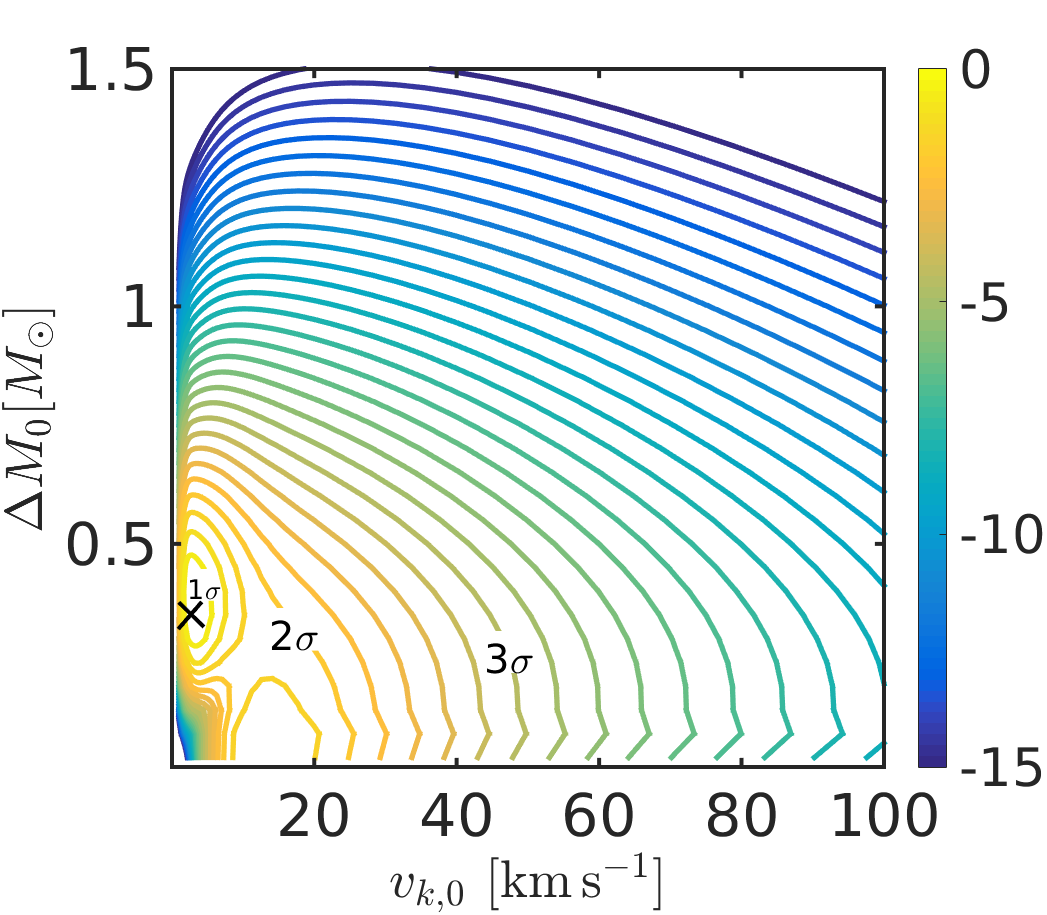}
\includegraphics[scale=0.24]{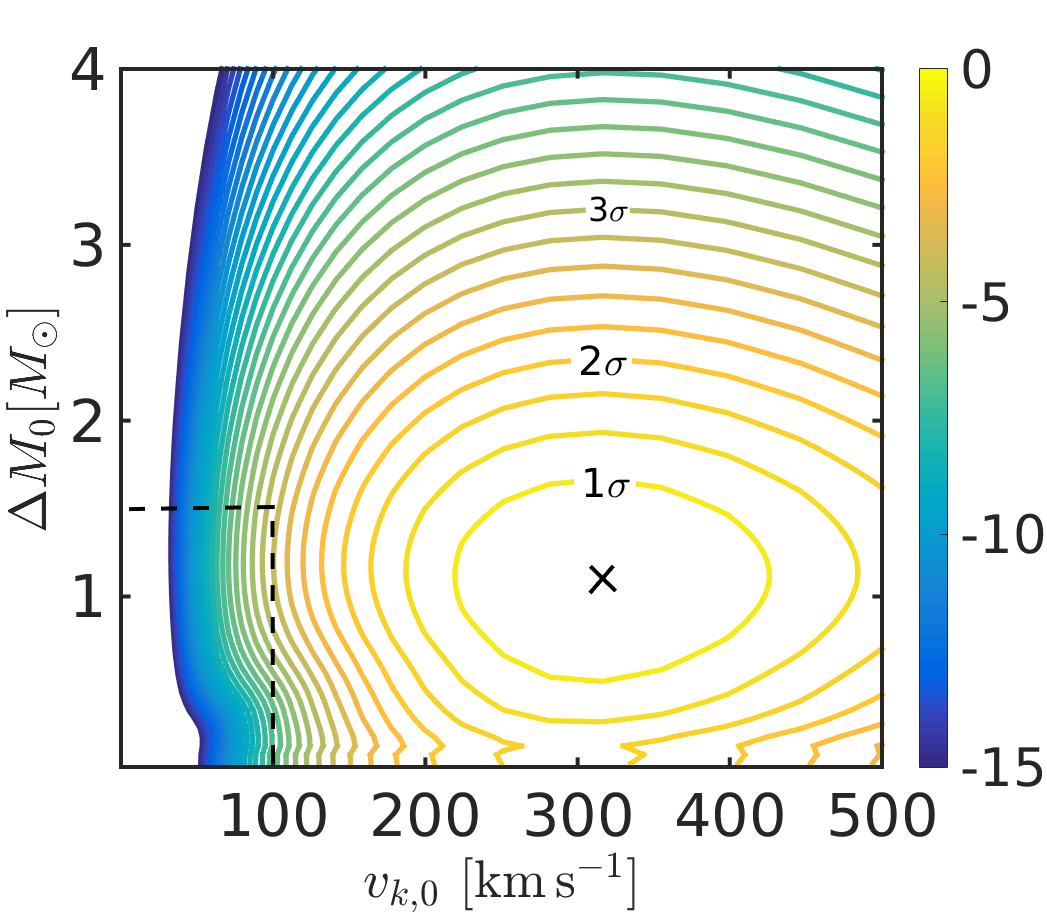}
\caption
{\small Likelihood functions as a function of $\Delta M_0,v_{k,0}$ for log-normal distributions in $\Delta M,v_k$ with $\sigma_M/ \Delta M_0=\sigma_{v_k}/v_{k,0}= 0.5$ assuming the orbital parameters at
$-t_{\rm max}$ (where $t_{\rm max}$ is the maximal possible age of the system, which in case it is unconstrained, is the minimum between
the time the system's eccentricity goes to zero and the Hubble time).Contour lines depict 0.5 logarithmic intervals of the likelihood function relative to the maximum value. Also shown are the $1\sigma$, $2\sigma$ and $3\sigma$ confidence intervals corresponding to $\ln L=-0.5,-2,-4.5$.
{\bf Left: }The 6 DNS systems categorized as ``small $e$ systems''. {\bf Right: }The 4 DNS systems
categorized as ``large $e$ systems''.
The dashed box in the right panel depicts the axis limits for the left figure.}
\label{fig:GausGausold}
\end{figure*}

We turn to explore the sensitivity of the results found above on the assumed parameters of the model.
First, we test the validity of using the present day orbital parameters in the calculations above. To this end we take the most conservative assumption, i.e. that the second neutron star was formed at a time
$t_{\rm max}$ ago, where $t_{\rm max}$ is the maximal possible age of the system
(with an exception for J0737-3039, J1906+0746 in which the regular pulsars provide a good estimate for the system's lifetime as mentioned in \S \ref{Sample}).
Most systems, have not evolved significantly during this time. For this reason we find that the best fit likelihood parameters
are within the error range of the results found in \S \ref{sec:bimodality}. A careful examination shows that in fact the two populations become even more distinct in this case, see Fig. \ref{fig:GausGausold}.

\begin{figure*}
\centering
\includegraphics[scale=0.24]{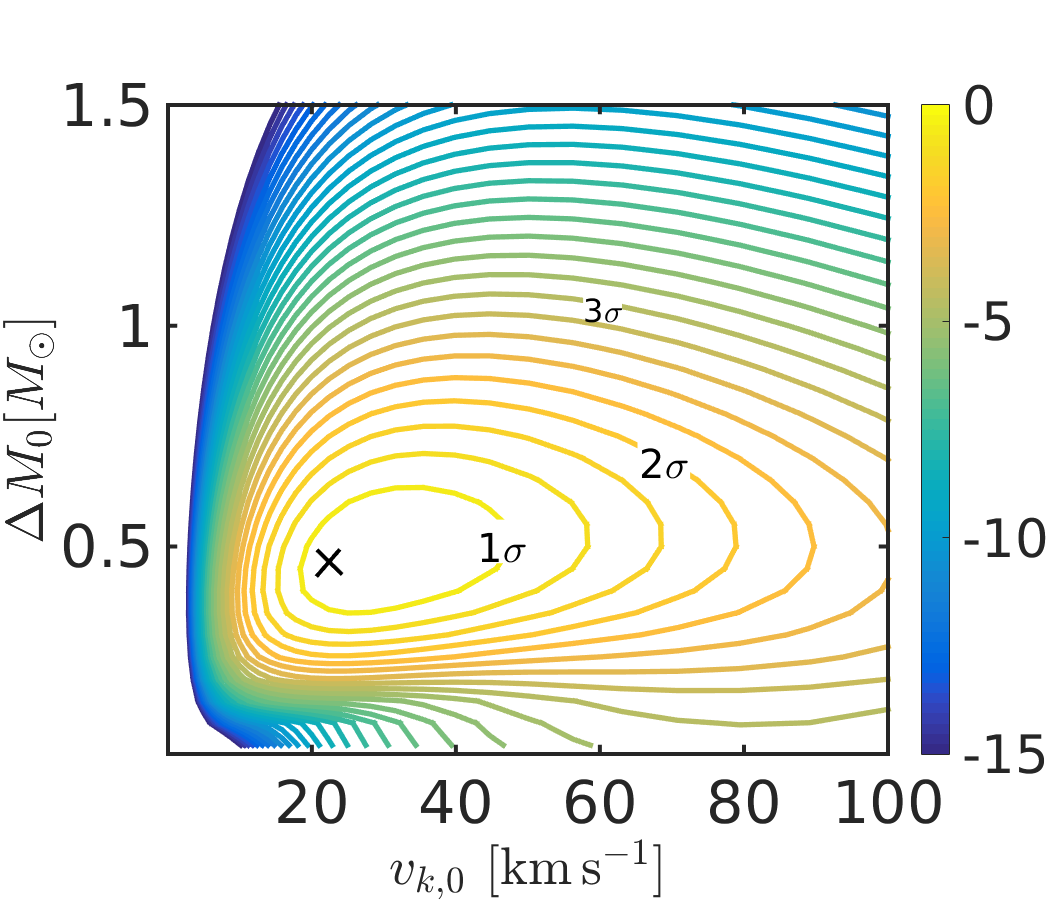}
\includegraphics[scale=0.24]{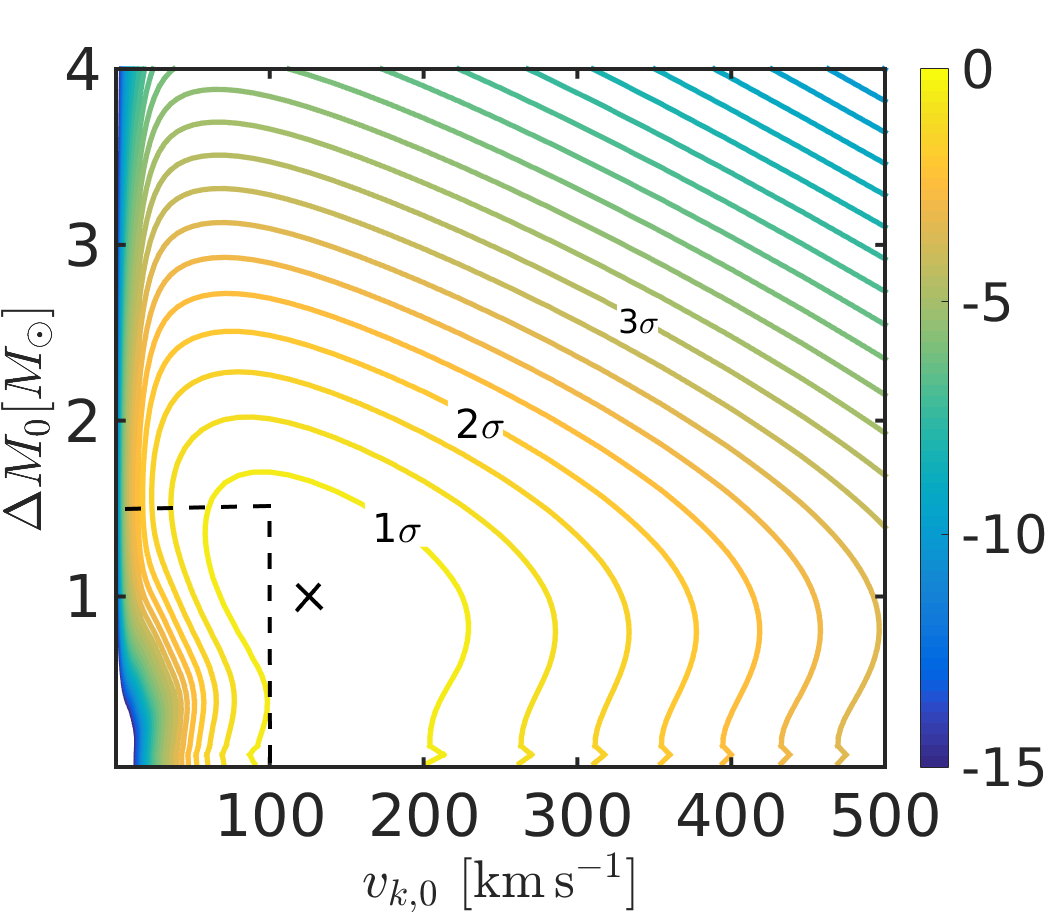}
\caption
{\small Likelihood functions as a function of $\Delta M_0,v_{k,0}$ for log-normal distributions in $\Delta M,v_k$ with $\sigma_M/ \Delta M_0=\sigma_{v_k}/v_{k,0}= 0.5$, and with B1534+12 grouped with the ``small $e$ systems''.
Contour lines depict 0.5 logarithmic intervals of the likelihood function relative to the maximum value. Also shown are the $1\sigma$, $2\sigma$ and $3\sigma$ confidence intervals corresponding to $\ln L=-0.5,-2,-4.5$.
{\bf Left: }The 7 DNS systems categorized as ``small $e$ systems''. {\bf Right: }The 3 DNS systems
categorized as ``large $e$ systems''. The dashed box in the right panel depicts the axis limits for the left figure.}
\label{fig:B1534}
\end{figure*}

Due to its large center of mass velocity (and spin tilt angle) we categorize B1534+12 as a ``large $e$ system'', although it has
marginal eccentricity and a relatively short spin period.
We carried out the likelihood analysis, this time with B1534+12 categorized as a
``small $e$ system''. This causes the peak of the mass ejection and kick velocity distributions for the ``small $e$ systems'' to slightly increase: $\Delta M_0^{s}=0.45M_{\odot}$ and $v_{k,0}^{s}=22 $km\,s$^{-1}$ (see Fig. \ref{fig:B1534}).
However, in this case too, the division to two groups is consistent with the data and results in a statistically significant improvement of the likelihood ratio test statistic as compared with the single group model.

\begin{figure*}
\centering
\includegraphics[scale=0.24]{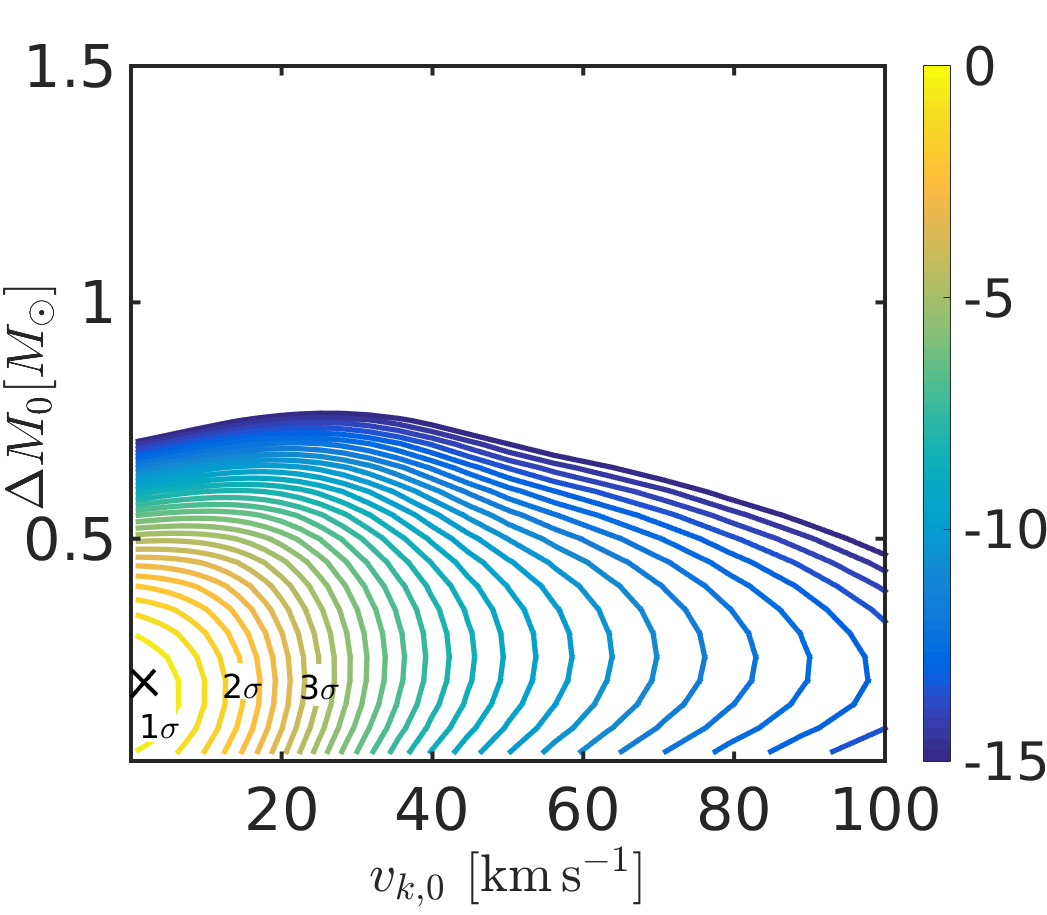}
\includegraphics[scale=0.24]{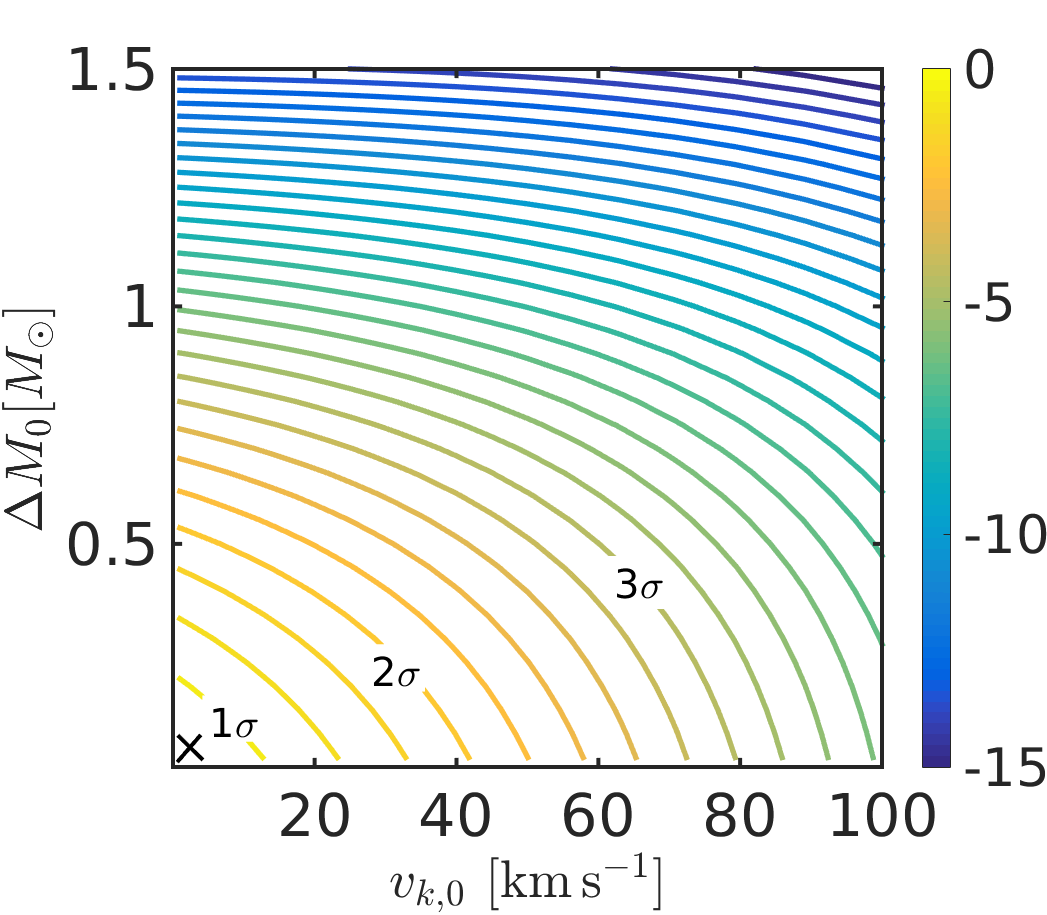} 
\caption
{\small Likelihood functions as a function of $\Delta M_0,v_{k,0}$ for Gaussian distributions of $\Delta M, v_k$ (limited to positive values only) and peaking at $\Delta M_0, v_{k_0}$ respectively,
for the DNS systems denoted as ``small $e$ systems''.
Both distributions have constant widths, independent of the peak value.
Contour lines depict 0.5 logarithmic intervals of the likelihood function relative to the maximum value. Also shown are the $1\sigma$, $2\sigma$ and $3\sigma$ confidence intervals corresponding to $\ln L=-0.5,-2,-4.5$.
{\bf Left: } narrow distributions in $\Delta M, v_k$, $\sigma_{\Delta M}=0.1M_{\odot}$ with $\sigma_{v_k}=10$km\,s$^{-1}$. $\Delta M_0^{s}>0.5M_{\odot}$ is strongly suppressed since
the average value of $e$ for these systems requires $\Delta M \approx 0.35M_{\odot}$ (see \S \ref{results}), and these values may not be reached for $\Delta M_0^{s}>0.5M_{\odot}$. 
{\bf Right: } wide distributions in $\Delta M, v_k$, $\sigma_{\Delta M}=0.5M_{\odot}$ with $\sigma_{v_k}=50$km\,s$^{-1}$. Due to the large ranges in $\Delta M, v_k$, the distributions are not well constrained in this case
and tend to the lowest possible values in both parameters.}
\label{fig:constwidth}
\end{figure*}

\begin{figure*}
\centering
\includegraphics[scale=0.24]{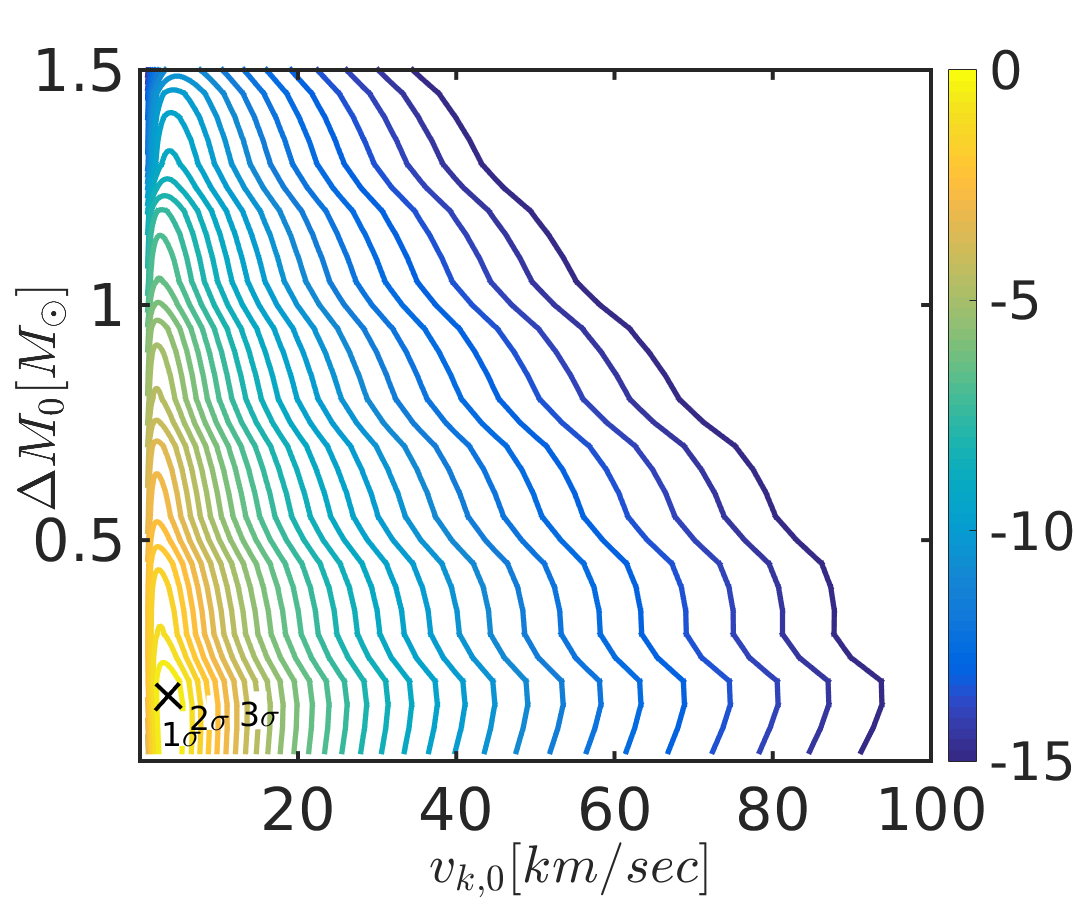}
\includegraphics[scale=0.24]{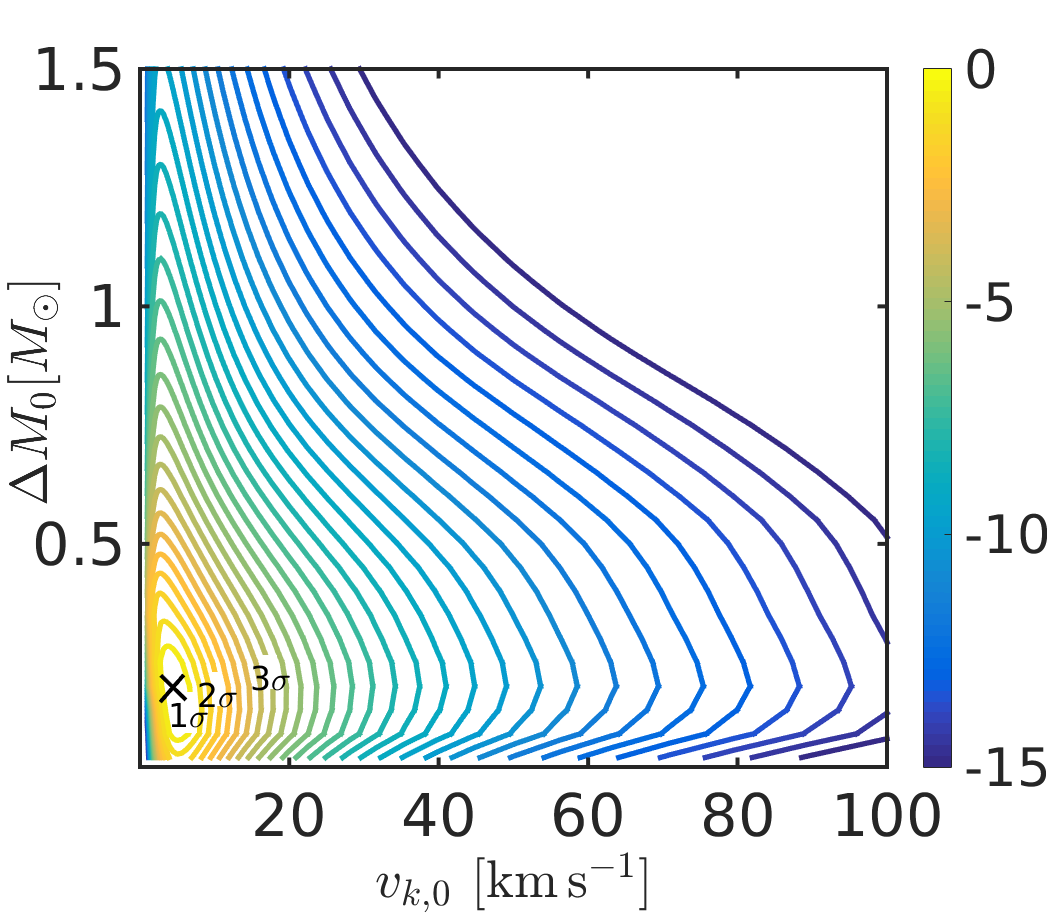} 
\includegraphics[scale=0.24]{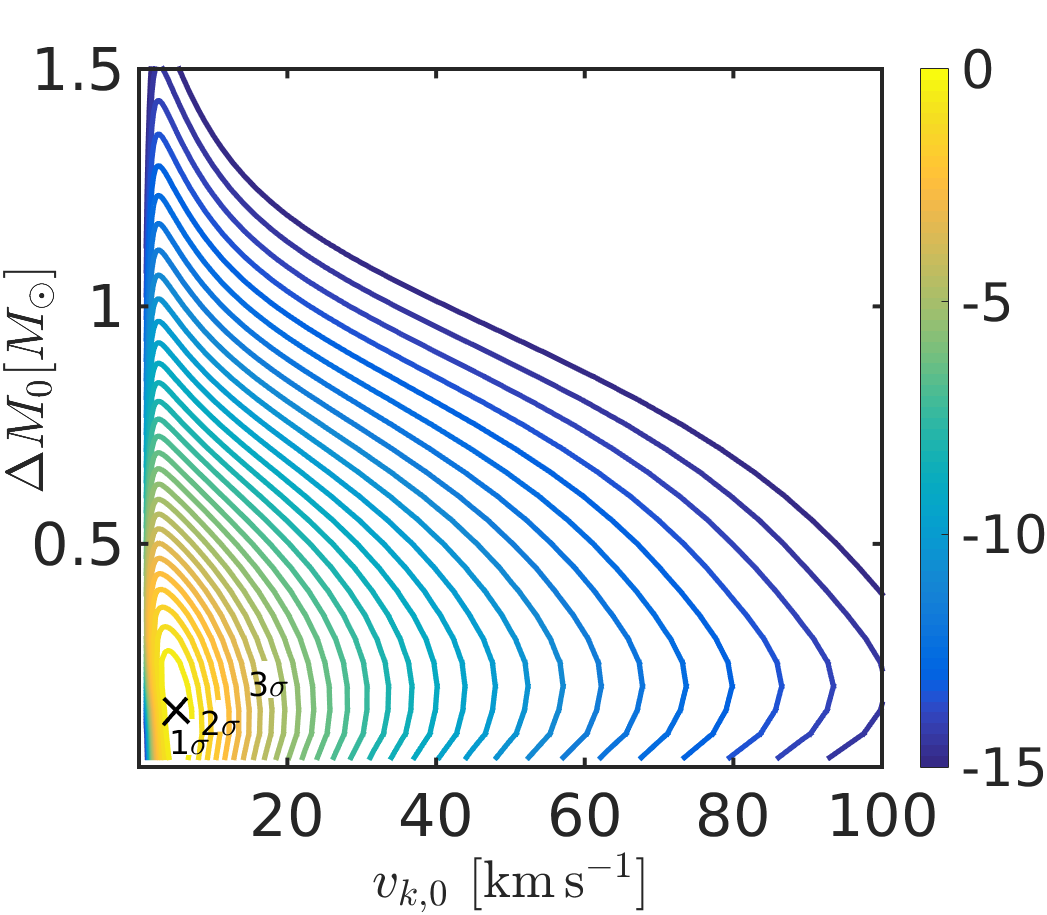} 
\includegraphics[scale=0.24]{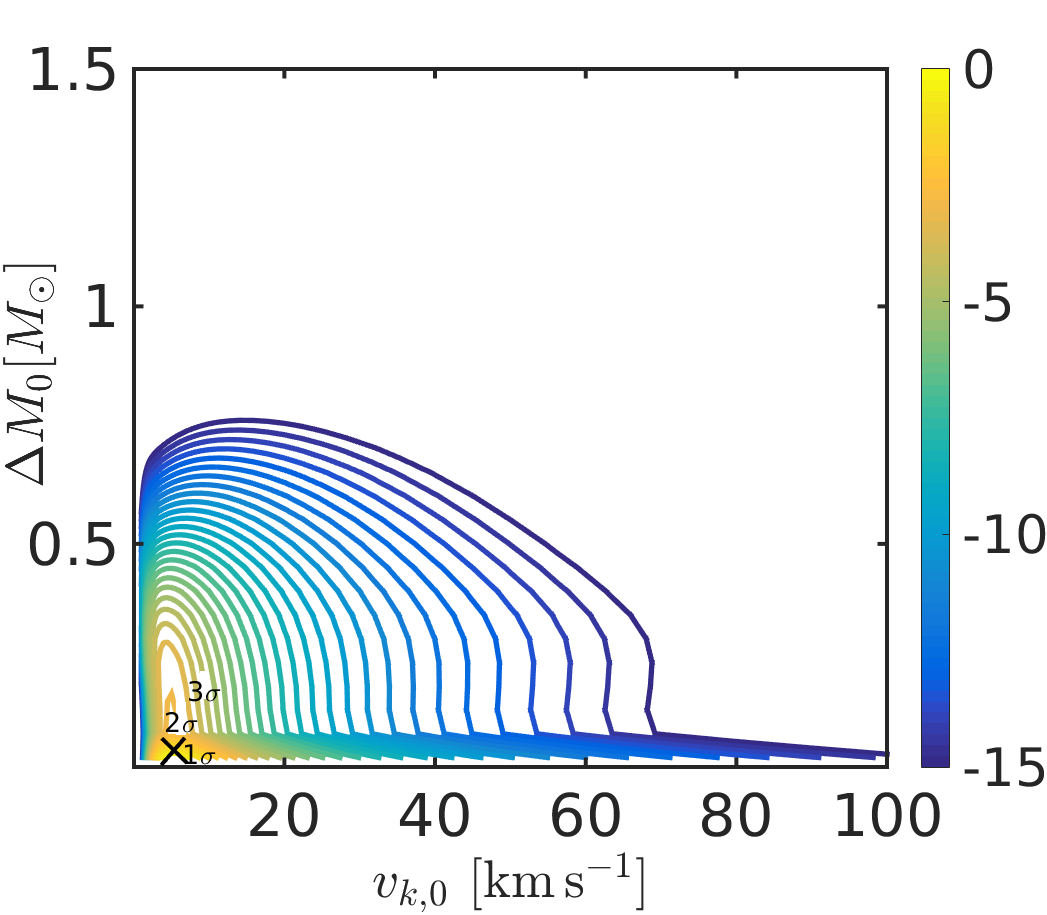} 
\caption
{\small Likelihood functions as a function of $\Delta M_0,v_{k,0}$ for different distributions of $\Delta M$ and for a log-normal
distribution in $v_k$ with an average of $v_{k,0}$ and a width $\sigma_{v_k}=v_{k,0}/2$ ($\sigma_{\ln v_k}=\sinh^{-1}(0.5 v_{k,0})$)
for the DNS systems denoted as ``small $e$ systems''. Contour lines depict 0.5 logarithmic intervals of the likelihood function relative to the maximum value. Also shown are the $1\sigma$, $2\sigma$ and $3\sigma$ confidence intervals corresponding to $\ln L=-0.5,-2,-4.5$.
{\bf Top Left: }An asymmetric distribution in $\Delta M$ - log-uniform in the range $[0.2\Delta M_0-2\Delta M_0]$.
{\bf Top Right: }A symmetric distribution in $\Delta M$ - uniform in the range $[0,2\Delta M_0]$.
{\bf Bottom: }Both panels depict Gaussian distributions of $\Delta M$, peaking at $\Delta M_0$ and with varying widths,
on the left with a standard deviation of $\sigma_{\Delta M}=0.5\Delta M_0$ and on the right with a standard deviation $\sigma_{\Delta M}=0.25\Delta M_0$.}
\label{fig:diffdist}
\end{figure*}

The largest uncertainty in our analysis is the shape of the distributions of $\Delta M$ and $v_k$ and the relation we have assumed between the width of the distribution and the average.
This latter relation is essential in order to limit the number of degrees of freedom, in view of the smallness of our sample.
In order to explore the model dependence on these assumed functional shapes we have carried the same analysis on a variety of different distributions: Gaussian (limited to positive values only), uniform, log-normal and log-uniform with varying  widths.
We find that in all cases, assuming two different populations results in a significant increase of the likelihood-ratio test-statistic, $LR$, compared to the single distribution model. In addition, we verify, using the KS test,
that that the eccentricity distributions for the two populations model are consistent with the observed one, whereas a single population is ruled out, at a $99\%$ confidence level.
The log-normal distributions presented earlier (with $\sigma_{\ln \Delta M}=\sigma_{\ln v_k}=\sinh^{-1}(0.5)$)
were chosen because they maximize the overall probability relative to the other functional shapes studied (this is done by comparing the maximum value of the likelihood functions obtained with different models).
As an illustrative example, we present here the results for different functional forms of the distributions for the likelihood maps of the population of  ``small $e$ systems''.
In all models we find $\Delta M_0^{s}=0.05-0.5M_{\odot}, v_{k,0}^{s}=3-30$km\,s$^{-1}$.
For a given width of the distributions, the fit is best for the log-normal distributions in mass ejections and kick velocities which are less symmetric (extending more to higher values than to lower values)
yet more peaked than the log-uniform distributions.
If  the typical widths of the distributions are independent of the average values, good fits are found with widths of $0.05-0.2M_{\odot}$ and $1-10$km\,s$^{-1}$ for the mass ejection and kick velocity respectively.
This is compatible with the values found for the best fit case described above, considering the location of the peak. In these cases, the range $\Delta M_0^{s}>0.5M_{\odot}$ becomes very strongly suppressed
since the average value of $e$ for these systems requires $\Delta M \approx 0.35M_{\odot}$ (see \S \ref{results}), and these values may not be reached for $\Delta M_0^{s}>0.5M_{\odot}$.
Results for these types of distributions are shown in Fig. \ref{fig:constwidth}.
On the contrary, if the widths of the distributions are proportional to the average values and the distributions are sufficiently narrow, regions with very small mass ejection ($\lesssim 0.2 M_{\odot}$) and kick velocity ($\lesssim 5$km\,s$^{-1}$) become excluded.
This is  essentially an artefact of these distribution that occurs because the smaller the peaks, the narrower the distributions become.
For sufficiently small average kick velocity and mass ejection, the solution with $\Delta M \approx 0.35M_{\odot}$ can no longer be probed by the distributions and the fits become very poor.
The likelihood analysis for distributions in which the width is proportional to the average value, does however favor relatively low $\Delta M_0$, since the best model is indeed one with quite a low mass ejection, and as one goes to lower values of $\Delta M_0$,
the probability density at the ``correct'' masses becomes larger.
Examples of different likelihood functions can be seen in Fig. \ref{fig:diffdist} for some representative cases of the functional forms considered.

\subsection{A single population with $\Delta M$ - $v_k$ correlation}

So far, we have assumed that the distributions of $\Delta M$ and $v_k$ are independent.
Alternatively, one should consider the possibility that the two are correlated. This happens naturally if the ``linear'' kick velocity
of the shell of ejected mass by the collapse
(this is the average velocity of the emitted material in one given direction)
 is constant. In this case:
\begin{equation}
 \Delta M=M_c \frac{v_k}{v_{\rm shell}} \ .
\end{equation}
We calculate the maximum likelihood for different values $v_{\rm shell}$ and different distributions for $v_k$. Now we carry out the analysis for the whole DNS population. 
The data is best fit with $v_{\rm shell}=100-200$km\,s$^{-1}$ and a log-normal distribution in $v_k$ with $\sigma_{\ln v_k}=\sinh^{-1}(2)$.
These values of $v_{\rm shell}$ seem reasonable, as typical velocities of the ejected material are likely to be around $few\times 1000$km\,s$^{-1}$, and an asymmetry
of order $10\%$ would then naturally result in ``linear'' kick velocities of this order.
In this case we find $v_{k,0}=15-25$km\,s$^{-1}$  corresponding to $\Delta M_0=0.14-0.23M_{\odot}$ (see Fig. \ref{fig:eProb}).
To test this model compared with the favoured model, with two separate populations, we run the likelihood ratio test.
We find a likelihood test statistic of $LR=18.4$, which, with a difference of two free parameters between the models, corresponds to a p-value: $p=0.02$, implying that this model cannot be ruled out statistically at the $3 \sigma$ level.
In addition, we find that the KS test for the eccentricity distribution, suggests  that this solution is acceptable. Thus the whole
DNS population can be fitted with by a single distribution if we assume a constant ejection velocity. 
We find similar consistent results also  for the other log-normal, uniform and log-uniform distribution functions.
While intriguing that we don't need two populations under this assumption
it should be stressed that the existence of such a correlation is consistent with the possibility of two different physical formation mechanisms,
one with small kick velocities and small mass ejection and another with larger kick velocities and larger mass ejection.

\begin{figure*}
\centering
\includegraphics[scale=0.2]{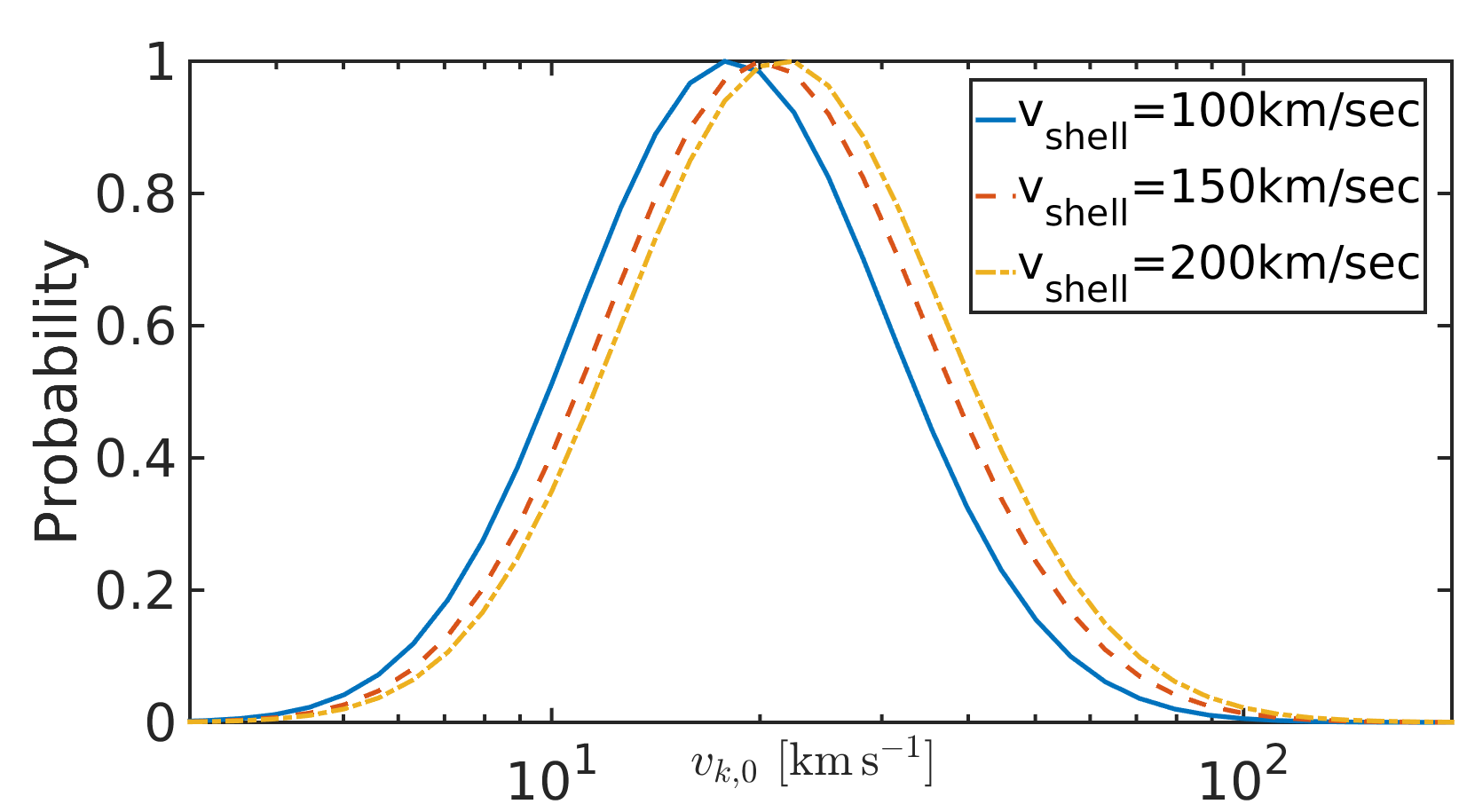}
\includegraphics[scale=0.2]{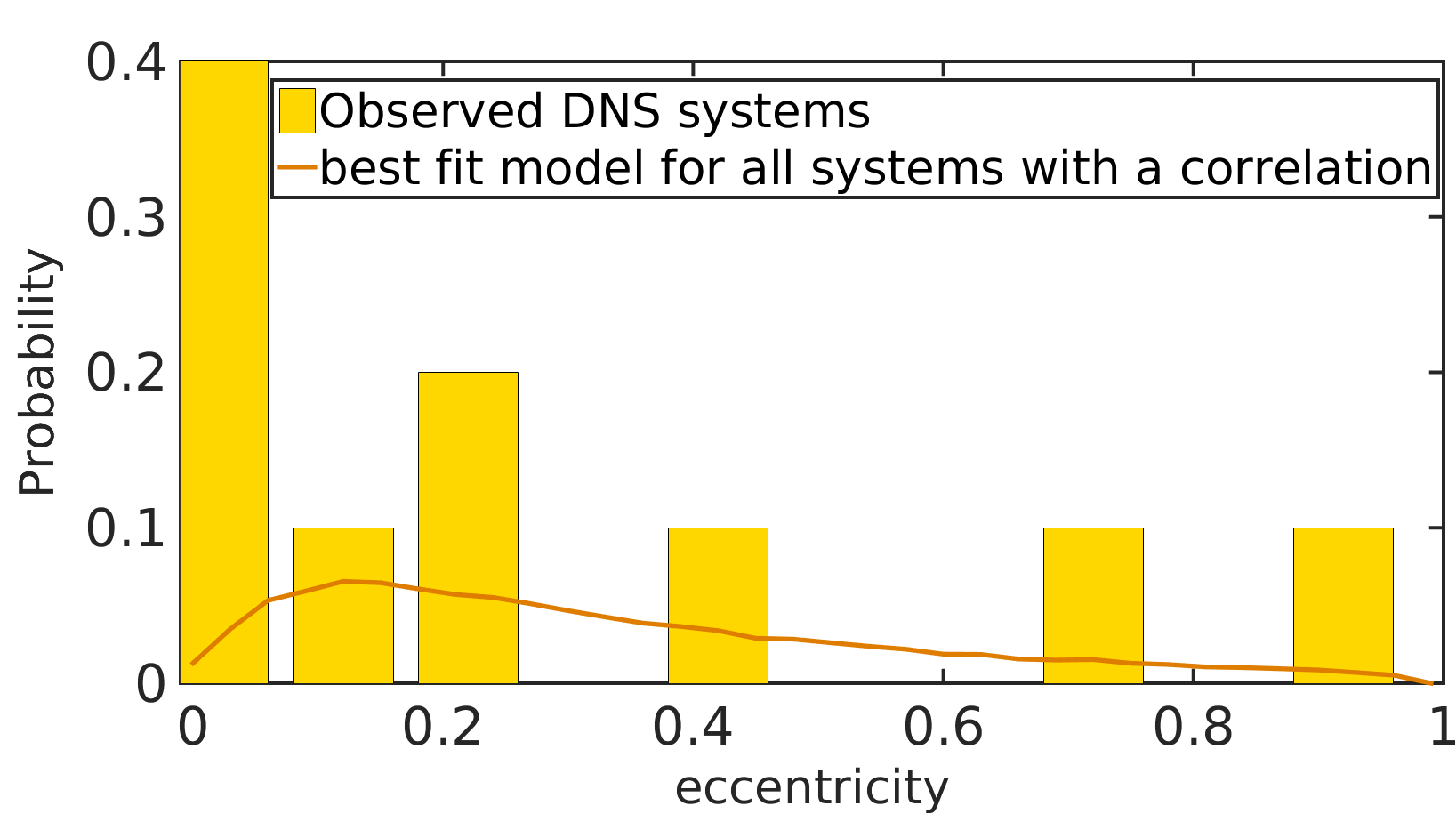}
\caption
{\small {\bf Left: } Likelihood function as a function of $v_{k,0}$ assuming a linear a correlation between $\Delta M$ and $v_k$ such that: $\Delta M=M_c \frac{v_k}{v_{\rm shell}}$ and a log-normal distribution in $v_k$
with an average of $v_{k,0}$ and a typical width $\sigma_{v_k}=2v_{k,0}$. {\bf Right: } Distribution of observed eccentricities as compared with the best fit model for the mass ejection and kick velocity
assuming a correlation between $\Delta M$ and $v_k$.}
\label{corr}
\end{figure*}

\subsection{Comparison with previous works}
\label{sec:compare}

\begin{figure*}
\centering
\includegraphics[scale=0.24]{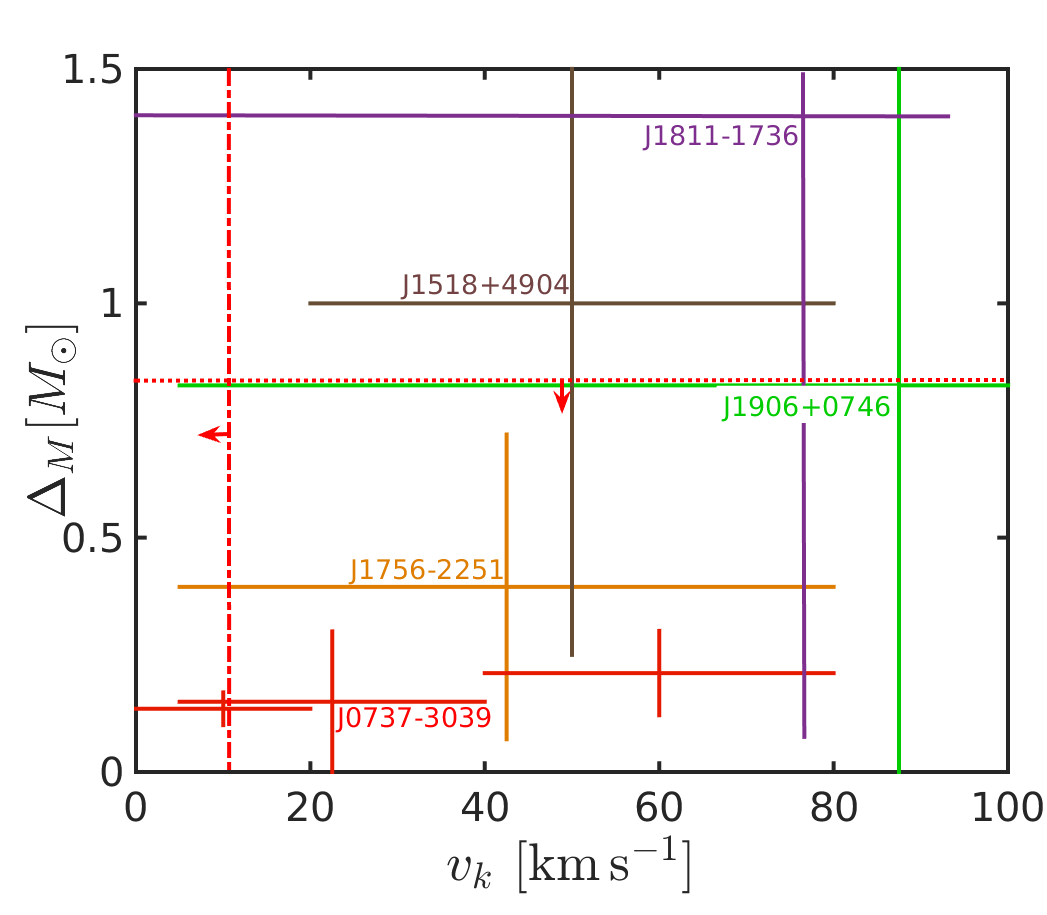}
\includegraphics[scale=0.24]{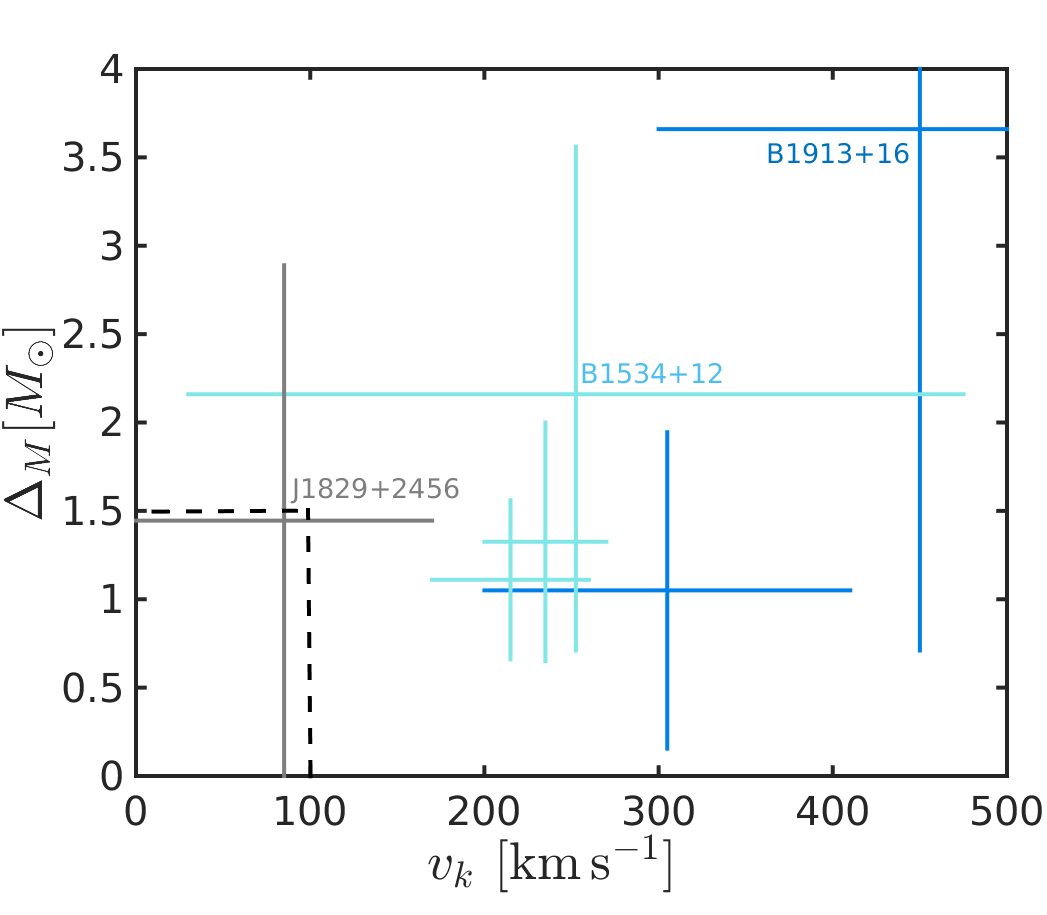}
\includegraphics[scale=0.24]{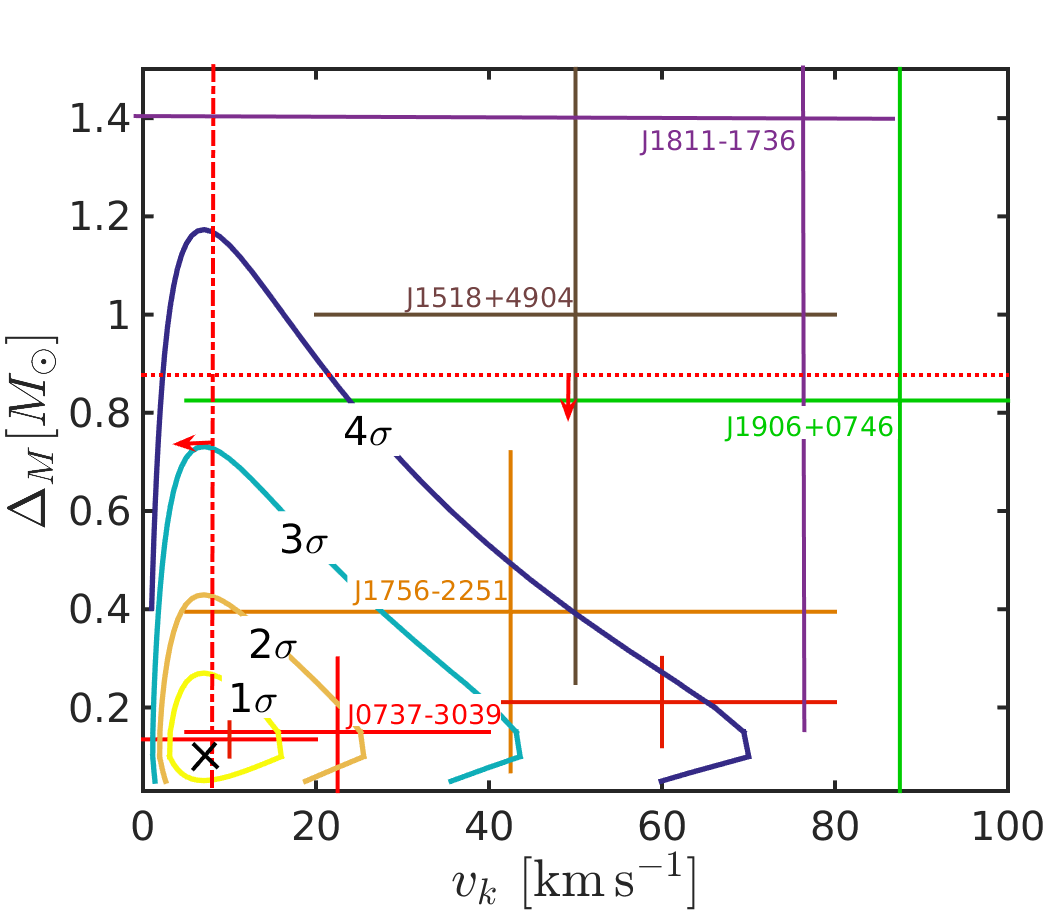}
\includegraphics[scale=0.24]{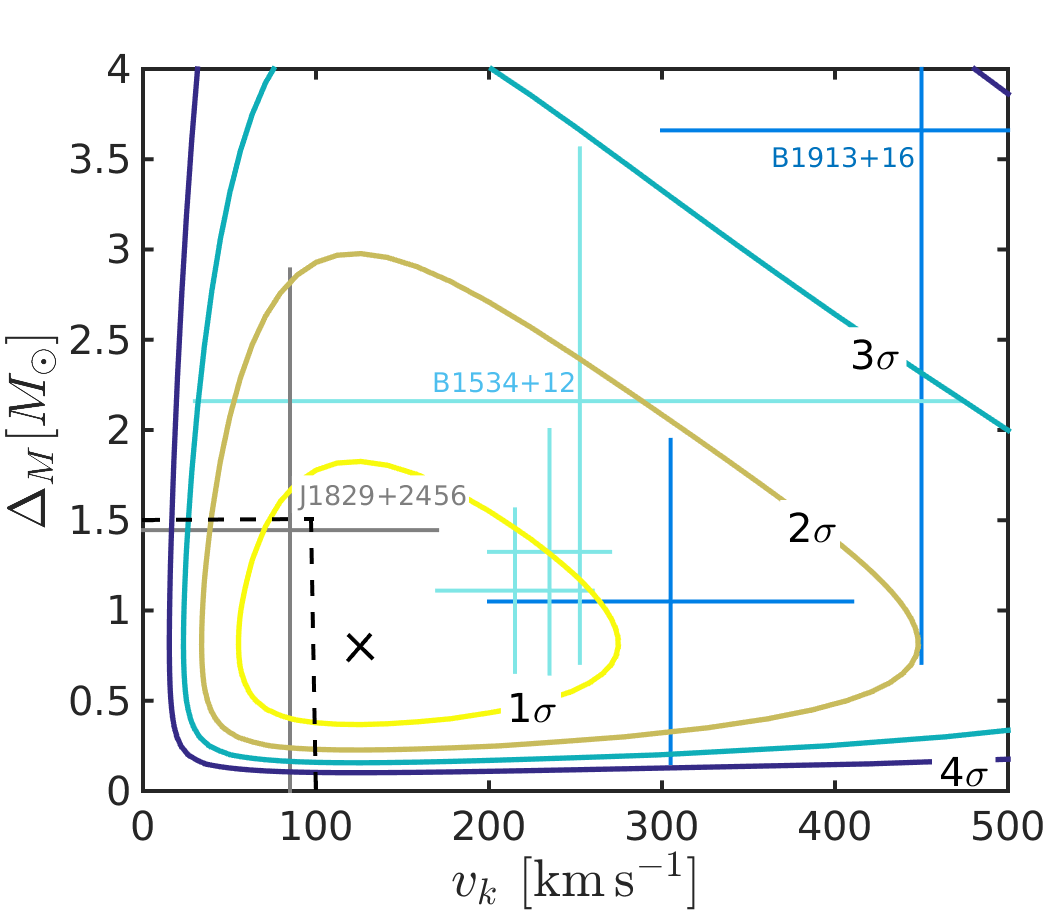}
\caption
{\small Constraints on the mass ejection and kick velocity obtained for some of the DNS systems in the sample by previous authors \citep{Kramer(2005),Stairs(2006),Willems(2006),Kalogera(2007),Wong(2010),Dallosso(2014)}. Notice that these pluses denote the 1$\sigma$ level limits. Generally the distributions found by these authors do not peak at the center points of these pluses.
{\bf Top Left: }DNS systems with small $e$ (J0737-3039 in red, J1756-2251 in orange, J1518+4904 in brown, J1811-1736 in purple and J1906+0746 in green). Two of the limits are upper limits only on either the kick velocity or the mass ejection (from \citealt{Kramer(2005)} and \citealt{Willems(2006)}) both are for J0737-3039.
They are denoted by dotted and dashed-dotted lines accordingly.
{\bf Top Right: }DNS systems with large $e$ (B1534+12 in teal, B1913+16 in blue and J1829-2456 in grey). The dashed box depicts the axis limits for the left figure.
{\bf Bottom: }Comparison between the likelihood distribution for the best fit parameters of mass ejecta and kick velocities found in our analysis
and the results in previous studies (on the left for systems with small $e$ and on the right for systems with large $e$).
The contour lines depict the $1,2,3$ and $4 \sigma$ levels for the likelihood distributions for the best fit parameters found in our analysis.
Although for most systems, our results are within the uncertainty limits of previous studies, there is a clear trend to lower kick velocities (and somewhat lower mass ejection) in our results (see \S \ref{sec:compare}). }
\label{fig:prev}
\end{figure*}

Earlier studies on the mass ejection and kick velocity during the formation of the second star in DNS have focused mostly on individual systems. 
We compare our results with 1$\sigma$ limits on the mass ejection and kick velocity obtained in these specific studies.
Fig. \ref{fig:prev} depicts a compilation of limits from previous studies, using various methods.
Notice that the center point of these limits are not necessarily the most likely mass ejections and kick velocities in the previous works. Specific distribution functions can be seen in detail in the corresponding papers.
We also overlay our best fit distributions obtained for those systems categorized as both ``small $e$ systems" and
``large $e$ systems" on top of our likelihood function for these groups of systems. {Although for most systems, the region with the highest likelihood (based on our results) is within the uncertainty limits of previous studies, there is a clear trend to lower kick velocities
(and somewhat lower mass ejection) in our results as compared with estimates by \cite{Wong(2010)} for J1518+4904, J1756-2251, J1811-1736 and J1906+0746 and with the estimates by \cite{Kalogera(2007)} for J0737-3039 and B1913+16. The main difference, we believe, arises from the fact that in their analysis \cite{Wong(2010)}  assume that $v_{kick}$ is uniformly distributed between 0 and $2500$km\,s$^{-1}$. 
This sets a prior on the velocity distribution that favours large kick velocities. \cite{Wong(2010)} continue calculating 
 possible mass transfer before the SN (due to Roche lobe overflow or stellar wind) and limit $M_{c,i}$ to be less than $8M_{\odot}$ (so as not to form a black hole). 
 Applying the observed constraints on the orbital parameters they find the probability distribution function for $V_{kick}$.
Similarly, \cite{Kalogera(2007)} also consider population synthesis models to account for mass transfer before the collapse. They also make an additional assumption setting a minimal pre-collapse mass $M_{c,i}$ which is assumed to be above $2.1M_{\odot}$ and therefore cannot allow for solutions with small mass ejection (found in the present study) by definition. 
These analyses are therefore different than ours in that they involve assumptions regarding a prior on the velocity distribution and the mass transfer process before the collapse and regarding the initial range of masses allowed to result in a neutron star - post collapse.
In the present work we constrain the kick and mass ejection directly from observations without using a population synthesis approach that relies on stellar evolution models.
Additionally, our analysis is different than previous works in that we are searching for probability distributions that fit multiple systems simultaneously, whereas in previous studies, each different system was treated individually.

\section{Discussion}
\label{discuss}
We have examined the evolution channels in DNS systems by constraining, using the orbital parameters of the systems today, the mass ejection and kick velocities received by the second NS during its formation.
Although the exact distribution of these parameters cannot be constrained due to the small number of available object, we show that it is impossible to account for the observed eccentricities of DNS systems given
independent distributions for the mass ejection and kick, which are the same for all systems. This is essentially due to the fact that there is a concentration of systems with low eccentricities along with several systems
with much larger eccentricities. We can account for the observed distribution if we assume that there are two different populations with two different distributions or if we assume that the mass and kick velocity are correlated, such that the ejected mass is always ejected with the same velocity.

The observed distribution is consistent with the assumption that systems with shorter pulsar spin periods are formed by a different channel than those with longer spin
periods. The former (``small $e$ systems") are constrained to have low mass ejections and weak kick velocities: $\Delta M_0\lesssim 0.5M_{\odot}$ and $v_{k,0}\lesssim 30 $km\,s$^{-1}$ (compatible, for example, with an electron capture supernova)
whereas the latter (``large $e$ systems'') can have mass ejections of up to $\sim 2.2 M_{\odot}$ and kick velocities of up to $400$km\,s$^{-1}$, in accord with regular Ib and Ic supernovae \citep{Drout(2011),Cano(2013),Bianco(2014)}.

Most (an exception is  J1906+0746, see \S \ref{Sample}) ``small $e$ systems" involve a rapid pulsar, which is most likely recycled and therefore it was the first born NS in the binary.
In particular, this may imply that the existence of a short spin period pulsar has affected 
the evolutionary trajectory of the companion or vice versa. This can happen, for example if
the two stars had rather low initial masses enabling them to be in a relatively close binary during the main sequence phase. Subsequently they were in a common envelope phase in which they lost most of their masses and finally  
after the first neutron star formed it was close enough to the companion to have
accreted a sufficient
amount of mass (of order $3-6\times 10^{-3} M_{\odot}$,  \citep{Dallosso(2014)}) to be mildly spun up as observed.
Then, by the time of the explosion, the companion has already lost so much mass, that its explosion is much weaker  \citep{MacLeod(2015)}.

Previous studies of the kick velocity distribution of pulsars: \citep{Lyne(1994),Hobbs(2005)} found a very high mean space velocity of regular single pulsars
with known proper motion. However, for msec pulsars, the velocities are significantly lower. 
\cite{Hobbs(2005)} studied 7 solitary msec pulsars and 28 msec pulsars in binaries and found 2D velocities of $77\pm16$km\,s$^{-1}$, $89\pm15$km\,s$^{-1}$ respectively.
In general, the change in the center of mass velocity is determined by two terms (see Eq. \ref{eq:vcm}), one due to the kick velocity, and the other due to the initial Keplerian velocity.
The observed center of mass velocities are of the same order as obtained from the Keplerian term in our models, and therefore these velocities are also consistent with models with kick velocities
significantly smaller than the Keplerian velocity (as indeed we find in our favoured models for the DNS systems).
Furthermore,  the observed ratio of $\sim 0.25$ between the number of isolated and binary systems with msec pulsars \footnote{In order to translate this to the intrinsic ratio, one has to take into account non-trivial observational effects regarding the probability of 
detecting an isolated vs binary msec pulsar} is within the range (0.2-0.5) of predicted ratios found in our best fit models for the ``small $e$ systems"
between  systems that are disrupted during the second collapse and those that remain bound. 
We thus speculate that it may be possible to extend the results presented here to the entire population of msec pulsars.

The observed orbital parameters of the whole DNS population are consistent with the existence of a correlation between the distributions of kick velocities and mass ejections. Specifically, if we assume that
the shell always receives the same speed due to the explosion, we may expect a linear correlation between the kick imparted to the NS and the amount of mass ejected. This model is consistent with the physical model described above and with the existence of
two groups of systems: those that receive weak kick velocities and small mass ejections and those that have stronger kick velocities and larger mass ejections.

Finally we mention that if indeed small $e$ systems in DNS are formed through small kick collapses (as is reflected by their relatively small proper motion velocities) then one may expect that most of these systems will reside close to the galactic disc.
On the other hand systems with large kicks will spend most of their time above the disk.
This may have important implications to the estimated merger rates of DNS and to the rate of LIGO/VIRGO detections. To quantify this effect, and in particular to estimate the intrinsic ratio of systems with ``small $e$" and small kicks to systems with
``large $e$" and large kicks one has to take into account a possible observational bias. 
To get an actual birth distribution, one needs to weight each system by $1/(V_{\rm max} T)$, where $V_{\rm max}$ is the total pulsar survey volume that
could have found each system, and $T$ is the system's lifetime.
Since the observed pulsars in both groups are recycled, and the dispersion in their periods is relatively small, the main factor that is likely to affect the observed fraction of systems belonging to each group is $V_{\rm max}$.
Systems with low kicks will spend most of the time close to the galactic plane while systems with larger kicks will spend significant amounts of time outside of the plane. Thus, if most pulsar searches are done close to the galactic
plane, this could introduce an observational bias in the ratio of the two groups. However, due to various high-latitude searches
for pulsars in recent years \citep{Burgay(2006),Jacoby(2009)}, this factor may not be so large. Further work is needed to fully understand the exact rates of the different groups. This, as well as the implications of these findings on the origin of short GRBs, on the rate of merges and on their time delay relative to the SFR will be discussed elsewhere.

{\bf Acknowledgements:} We thank S. Phinney, K. Hotokezaka, V. M. Kaspi and M. Modjaz for helpful discussions. We also thank the Israeli I-core centre. This research was supported by a CNSF-ISF grant and an ISA grant.

\end{document}